\documentclass[aps,twocolumn,prl,showpacs,preprintnumbers,superscriptaddress,amsmath,amssymb,floatfix]{revtex4}
\usepackage[charter]{mathdesign}	
\usepackage{amsmath}	
\usepackage{graphicx}
\usepackage{dcolumn}
\usepackage{bm}
\usepackage[english]{babel}
\usepackage{verbatim}
\usepackage{color}
\usepackage[percent]{overpic}

\usepackage[normalem]{ulem}
\newcommand{\ms}[1]{\textcolor{cyan}{#1}}
\newcommand{\jm}[1]{\textcolor{blue}{#1}}

\begin{document}
\preprint{APS/123-QED}

\author{M. Schneider}
\affiliation{I. Physikalisches Institut,
Georg-August-Universit\"{a}t G\"ottingen, D-37077 G\"ottingen}
\author{D. Geiger}\thanks{present address: Institut f\"ur Festk\"orperphysik, Technische Universit\"at Wien}
\affiliation{1. Physikalisches Institut, Universit\"at Stuttgart, D-70550 Stuttgart, Germany}
\author{S. Esser}
\affiliation{I. Physikalisches Institut,
Georg-August-Universit\"{a}t G\"ottingen, D-37077 G\"ottingen}
\author{U. S. Pracht}
\affiliation{1. Physikalisches Institut, Universit\"at Stuttgart, D-70550 Stuttgart, Germany}
\author{C. Stingl}
\affiliation{I. Physikalisches Institut,
Georg-August-Universit\"{a}t G\"ottingen, D-37077 G\"ottingen}
\author{Y. Tokiwa}
\affiliation{I. Physikalisches Institut,
Georg-August-Universit\"{a}t G\"ottingen, D-37077 G\"ottingen}
\author{V. Moshnyaga}
\affiliation{I. Physikalisches Institut,
Georg-August-Universit\"{a}t G\"ottingen, D-37077 G\"ottingen}
\author{I. Sheikin}
\affiliation{Laboratoire Nationale des Champs Magn\'etiques Intenses, 25, Rue des Martyrs,
38042 Grenoble, France} 
\author{J. Mravlje}
\affiliation{Josef Stefan Institute, SI-1000, Ljubljana, Slovenia}
\author{M. Scheffler}
\affiliation{1. Physikalisches Institut, Universit\"at Stuttgart, D-70550 Stuttgart, Germany}
\author{P. Gegenwart}
\affiliation{I. Physikalisches Institut,
Georg-August-Universit\"{a}t G\"ottingen, D-37077 G\"ottingen}

\title{Low-energy electronic properties of clean CaRuO$_3$: elusive Landau quasiparticles}

\date{
\today%
}

\begin{abstract}
We have prepared high-quality epitaxial thin films of CaRuO$_3$ with residual resistivity ratios up to 55. Shubnikov-de Haas oscillations in the magnetoresistance and a $T^2$ temperature dependence in the electrical resistivity only below 1.5~K, whose coefficient is substantially suppressed in large magnetic fields, establish CaRuO$_3$ as a Fermi liquid (FL) with anomalously low coherence scale. Non-Fermi liquid (NFL) $T^{3/2}$ dependence is found between 2 and 25~K. The high sample quality allows access to the intrinsic electronic properties via THz spectroscopy. For frequencies below 0.6 THz, the conductivity is Drude-like and can be modeled by FL concepts, while for higher frequencies non-Drude behavior, inconsistent with FL predictions, is found. This establishes CaRuO$_3$ as a prime example of optical NFL behavior in the THz range.
\end{abstract}


\maketitle Landau Fermi liquid (FL) theory assumes a one-to-one correspondence of electronic excitations in metals (so-called quasiparticles, QPs) to those of the non-interacting Fermi
gas. Remarkably, FL behavior persists in materials where the electronic interactions are strong, often, however, only below a low FL temperature $T_\mathrm{FL}$.  FL theory is the established framework that describes metallic behavior in a broad class of materials ranging from elemental metals to correlated oxides and heavy-fermion systems. Therefore metals that behave inconsistently with FL predictions attract a lot of attention.  A quasi-linear instead of FL-predicted quadratic temperature dependence of the electrical resistivity is found in heavy-fermion metals near a magnetic quantum critical point (QCP) and in high-temperature superconductors at optimal doping ~\cite{Cooper,Kasahara,Mathur}. It is often controversially discussed whether such non-Fermi-Liquid (NFL) behavior arises due to the interaction of the QPs with low-energy magnetic excitations near QCPs or the QP picture needs to be abandoned completely~\cite{vL}.

NFL behavior has also been discussed in certain ruthenates, i.e.\ 4d oxides that are, unlike the heavy-fermion compounds and cuprates, characterized by broad electronic bands (with bandwidth $W\sim 3$~eV) and small interaction strength $U\lesssim W$. The optical conductivity of SrRuO$_3$ and CaRuO$_3$ in the infrared exhibits an unusual frequency dependence $\sigma_1 \propto \omega^{-0.5}$~\cite{Kostic1998,Lee2002}, while time-domain THz measurements found signs of an unconventional metallic response at lower frequency~\cite{Kamal2006}. However, disorder strongly influences the properties of ruthenates~\cite{Capogna} and FL behavior was observed in sufficiently clean samples of SrRuO$_3$. CaRuO$_3$ was argued to be positioned right at a magnetic QCP, as signified by a logarithmic term in the specific heat coefficient and an electrical resistivity $(\rho(T)-\rho_0) \propto T^{3/2}$ ($\rho_0$: residual resistivity) below 30~K~\cite{Capogna,Cao08}. Since paramagnetic CaRuO$_3$ crystallizes in the same orthorhombically distorted perovskite structure as the isoelectronic itinerant ferromagnet SrRuO$_3$ ~\cite{Longo,Cao}, proximity to a ferromagnetic instability indeed seems plausible. The substitution of Sr by Ca in Sr$_{1-x}$Ca$_x$RuO$_3$ results in a suppression of the ordering beyond $x\approx 0.7$~\cite{Yoshimura,Khalifah,Cao,Schneider}. This is associated with a drastic softening of magnetic
exchange~\cite{Mazin,Srimanta}. First-principles band-structure calculations indicate that CaRuO$_3$ is located close to a ferromagnetic instability and paramagnon-like spin excitations are very soft~\cite{Mazin}. Ferromagnetic dynamical scaling has been found in NMR experiments although the bulk susceptibility reveals a strongly negative Curie-Weiss temperature in CaRuO$_3$~\cite{Yoshimura}.

Complementary insights into ruthenates have come from the dynamical mean-field theory (for a review, see Ref.~\cite{Georges2013}) which pointed out how the Hund's rule coupling suppresses the coherence scale and induces strong electronic correlations even in cases where
$U\lesssim W$.  Interestingly, even in such a local picture, NFL effects are found, with ongoing debate about how much can they persist into the zero-temperature limit ~\cite{Werner2008,Georges2013, Yin2012}.
The material that might realize such a local NFL is CaRuO$_3$.
 Precise data on CaRuO$_3$ is thus strongly desired but difficult to obtain, as in CaRuO$_3$ the crystal growth is found to be sophisticated~\cite{Cao,Kikugawa} and does not lead to pure-enough samples to address whether the $T^{3/2}$ behavior is a consequence of disorder or intrinsic and whether at low-enough temperatures a FL ground state is established. Answering these questions demands improved quality of the samples. This challenging task is worth undertaking also because other pervoskite ruthenates like Sr$_2$RuO$_4$~\cite{Maeno} or Sr$_3$Ru$_2$O$_7$~\cite{Sr327review} have shown fascinating low-temperature states in extremely clean samples.

Previously, thin films of CaRuO$_3$ have been synthesized e.g. by sputtering and pulsed-laser deposition
techniques~\cite{Eom,Klein,Khalifah,Wissinger}. However, residual resistivity ratio (RRR) values of order 5-10 only have been obtained. Below, we report the synthesis of high-purity CaRuO$_3$ thin films, first observation of quantum oscillations, and a study of the mass enhancement in this material. Our results demonstrate a fragile FL ground state and robust NFL behavior in its vicinity at elevated temperatures. The coherence scale $T_\mathrm{FL}$ increases rapidly under applied magnetic field, signaling a strong influence of spin fluctuations on the electronic properties. Low-temperature THz conductivity data indicate that at frequencies above 0.6 THz the optical response is inconsistent with simple Drude behavior and FL predictions. Instead, we find a very strong frequency dependence of the optical scattering rate, which highlights the role of electronic interactions also for the optical properties of CaRuO$_3$.

\begin{figure}
\includegraphics[width=0.95\linewidth]{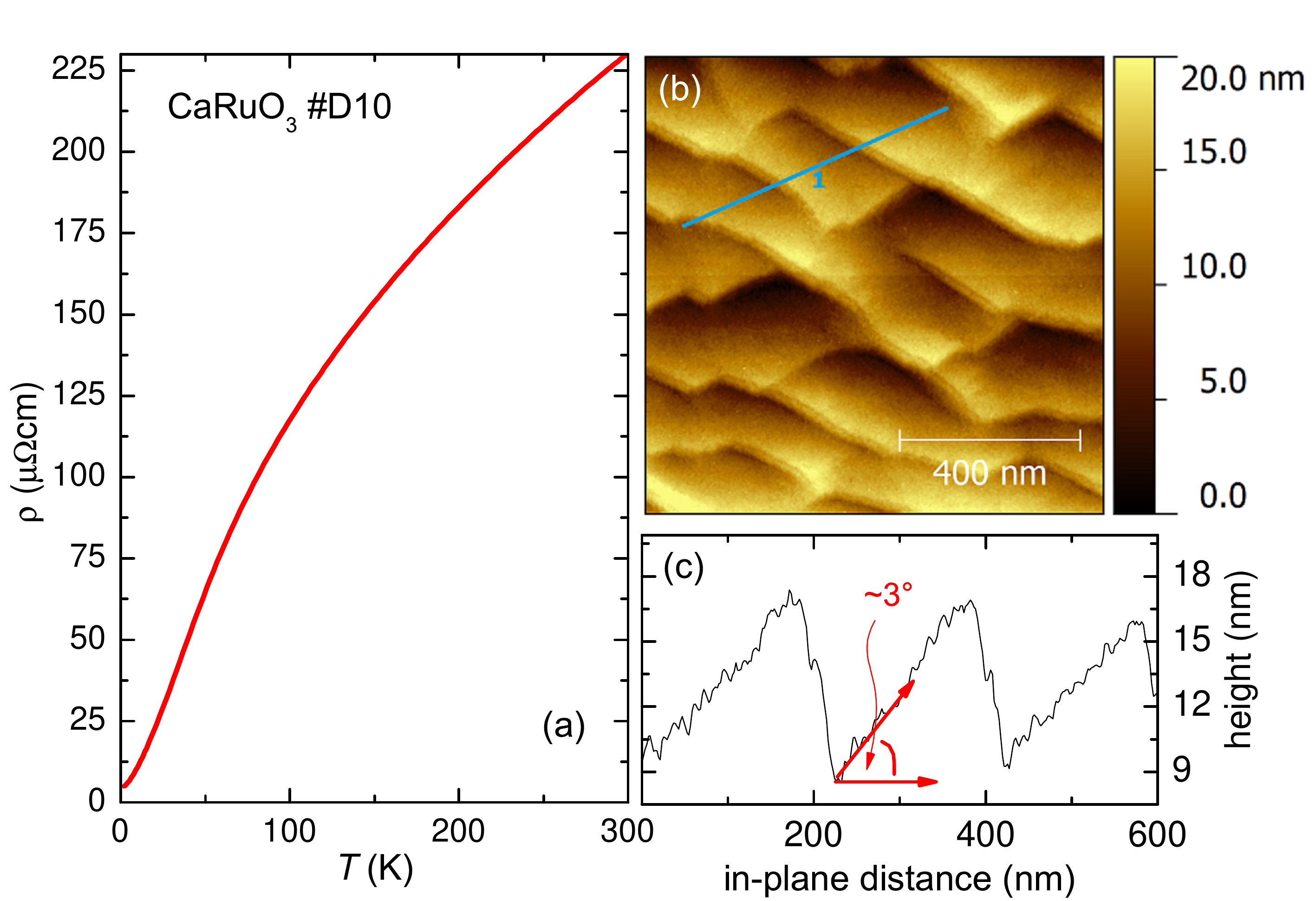}
\caption{(color online) (a) Temperature dependence of the electrical resistivity of a 77 nm thin film of CaRuO$_3$. (b) topography of a $(1\times 1)\mu$m$^2$ square as determined by room temperature STM. (c) Corresponding height profile along the direction indicated by the blue line from right to left in part b.}
\label{Fig1}
\end{figure}

Details of the CaRuO$_3$ thin film synthesis by the metalorganic aerosol deposition technique, as well as their characterization are provided in Supplementary Material (SM)~\cite{SM}. An important improvement of the sample quality has been obtained by using (110) oriented NdGaO$_3$ substrates with $3^\circ$ miscut angle, yielding RRR values up to 55 which is comparable to the best SrRuO$_3$ thin films~\cite{MackenzieSRO}. Fig. 1a displays the temperature dependence of the electrical resistivity of a CaRuO$_3$ thin film. The surface morphology shown by a scanning tunneling microscopy (STM) image (Fig. 1b) clearly indicates a step bunching growth mode with steps of about 10 nm height and 220 nm lateral extension, resulting from the $3^\circ$ miscut angle of the NGO substrate (cf. the profile in Fig. 1c). This evidences a perfect crystal growth. The preferential orientation of the steps is compatible with in-plane epitaxial growth, while out-of-plane epitaxy is also evidenced by XRD (see SM). High-resolution transmission electron microscopy measurements (see SM) indicate homogeneous growth of CaRuO$_3$ without any indication for secondary phases or grain boundaries.

\begin{figure}
\includegraphics[width=0.95\linewidth]{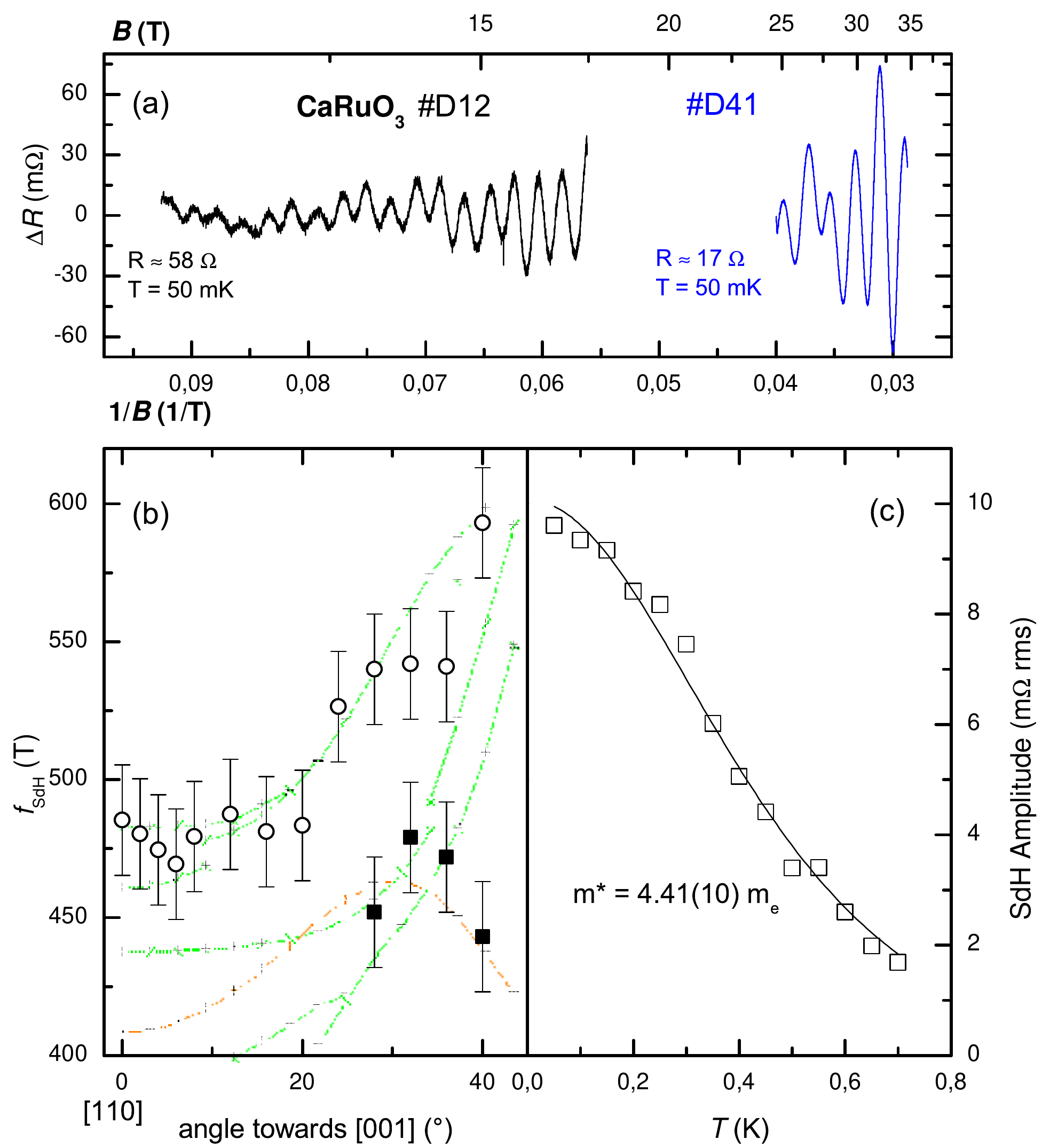}
\caption{(color online) (a) Shubnikov-de Haas (SdH) oscillations as obtained by subtraction of a second-order polynominal from the isothermal magnetoresistance of CaRuO$_3$ at 50~mK for two different thin films at two different field ranges for $B\parallel (110)$. (b) Angular dependence of the oscillation frequencies (symbols) compared to the LDA prediction (lines). (c) Temperature dependence of the SdH amplitude for sample D12. Solid line indicates fit to the Lifshitz-Kosevich formula yielding an effective mass of $4.4m_e$.} \label{Fig2}
\end{figure}

The excellent quality of our thin films is corroborated by the  observation of Shubnikov-de Haas (SdH) oscillations in the low-temperature magnetoresistance of CaRuO$_3$. Measurements on two different samples, up to 17 T in a superconducting solenoid and between 25 and 35 T in a high-field electromagnet (see also SM) reveal consistently quantum oscillations with a frequency of 470~T, corresponding to a Fermi surface area of $4.5\times 10^{14}$ 1/cm$^2$ and $k_{\rm F}=1.2 \times 10^9$m$^{-1}$. The temperature dependence of the SdH amplitude follows the expected Lifshitz-Kosevich behavior with an effective mass $m^\star =4.4m_e$ (Fig. 2c). Simultaneous observation of a quadratic resistivity below 1.5~K (discussed below) proves that the material is a Fermi liquid, which overrules pictures explaining the unusual properties in terms of the NFL ground state. 

We computed the fermiology of CaRuO$_3$ with the density functional theory within the local-density approximation (LDA) (see SM). LDA reveals many Fermi surface sheets that cover the
frequency range between 0.1 and 1 kT quite densely with some
additional long orbits extending (depending on the field direction) up to 10~kT. While the observed data (Fig. 2b) could only be fitted assuming several separate frequencies, the
limited number of oscillations does not allow their precise
determination. In addition, rotating the field out of the direction
perpendicular to the thin film plane results in a decrease of the SdH signal and disappearance beyond about $40^\circ$. Comparison with LDA band structure suggests that the observed two frequency branches belong to extremal orbits of Fermi surface sheets $\delta_2$ and $\beta_3$ (see SM). The LDA masses of both orbits are about $1.0~m_e$. The observed mass enhancements due to correlations, $\sim 4.4$ and $\sim 4.2$ for the former and latter orbit, respectively, are thus a bit smaller than the average enhancement $\sim 7.0$ at zero field estimated from the experimental specific heat coefficient of 73~mJ/molK$^2$~\cite{shepard97}. Interestingly, assuming a Kadowaki-Woods relation $m^\star\propto \gamma\propto A^{1/2}$, the reduction of $A$ by a factor 2 between 0 and 16 T (see below) accounts for most of the difference between the "specific heat mass" and the effective mass observed by SdH oscillations. The field dependence of the oscillation amplitude (not shown) yields an electronic mean free path of 40 nm characteristic for a clean metal, and a quasiparticle scattering time of $\tau_\mathrm{qp}\sim1.3$~ps. Using the transport life time $\tau_\mathrm{transp}=2Z \tau_\mathrm{qp}$ (taking for $Z$ the inverse mass enhancement), the conductivity evaluated from the band theory is $\sim 4\mu \Omega$cm, consistent with the residual resistivity of our samples.

\begin{figure}
\includegraphics[width=1.0\linewidth]{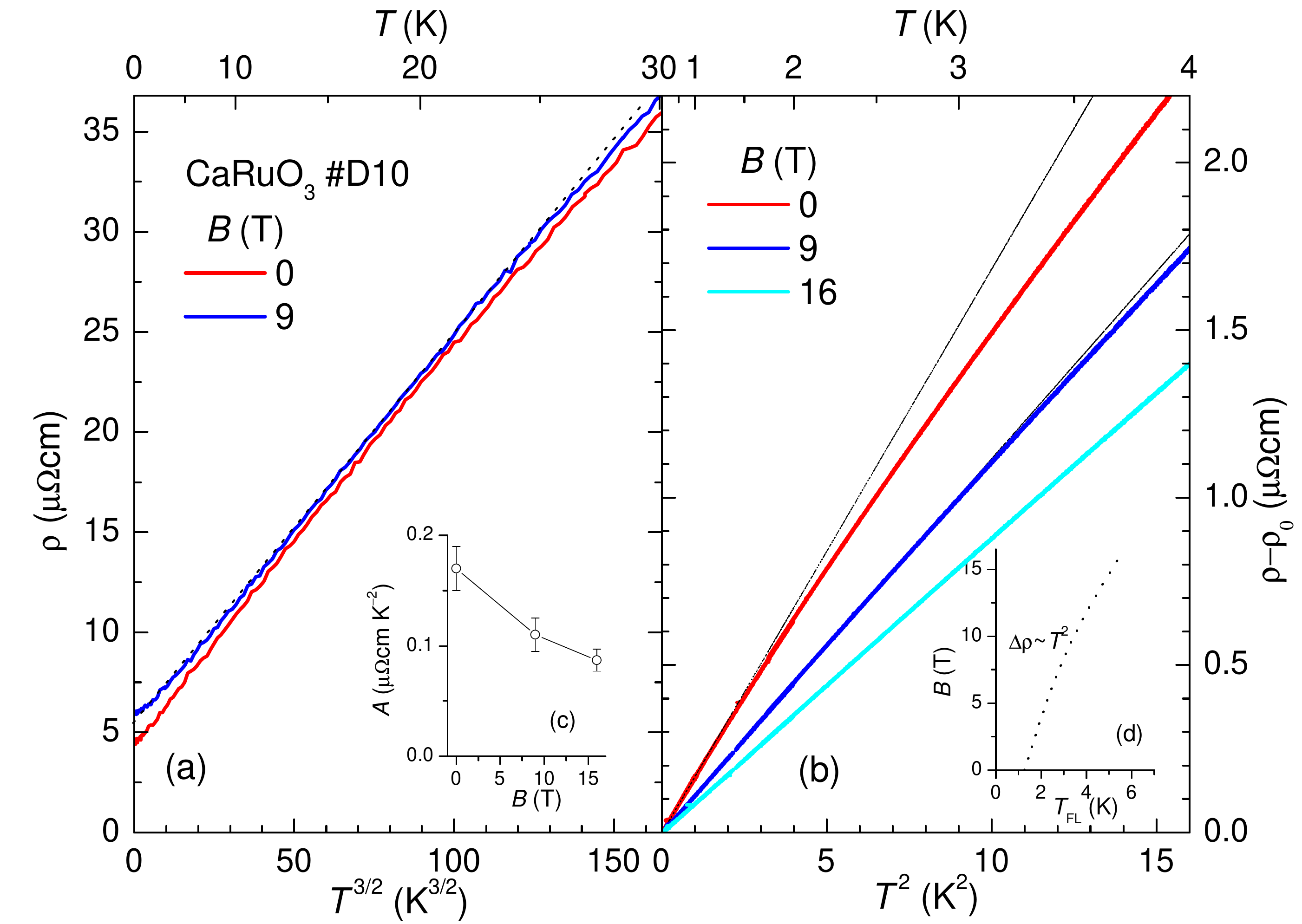}
\caption{(color online) (a) Electrical resistivity of CaRuO$_3$ vs. $T^{3/2}$ at  zero field and a field of 9~T applied transverse to the film plane. Dotted straight line indicates $T^{3/2}$ dependence. (b) Low-temperature (between 50 mK and 4~K) electrical resistivity as $\rho-\rho_0(B)$ vs $T^2$ for various fields. Straight dotted lines indicate Fermi liquid behavior. (c) Coefficient $A$ from $\rho-\rho_0=AT^2$ behavior vs. magnetic field. (d) Phase diagram indicating the existence range of Fermi liquid $T^2$ behavior in the electrical resistivity.}
\label{Fig4}
\end{figure}

In Fig. 3b, we show the temperature-dependent part of the electrical resistivity $\Delta\rho(T)=\rho(T,B)-\rho_0(B)$ down to 50~mK. At low $T$, the data follow $\Delta\rho(T)=A(B)T^2$. However, this FL behavior is limited to temperatures below a tiny $T_\mathrm{FL}=1.5$~K. At zero magnetic field, $T_\mathrm{FL}$ (cf. Fig. 3d) is not only small compared to other oxides (e.g., vanadates and molybdates with $T_\mathrm{FL}>100$~K) but is also substantially smaller than $T_\mathrm{FL}=7$~K found in Sr$_3$Ru$_2$O$_7$~\cite{Capogna}. Taking into account that the strength of correlations as suggested by the specific heat enhancement in Sr$_3$Ru$_2$O$_7$ is actually larger (about 10, at zero field), the FL behavior in CaRuO$_3$ is surprisingly fragile.

The low coherence scale $T_\mathrm{FL}$ increases strongly in the magnetic field, while the coefficient $A$ is reduced by a factor 2 between 0 and 16 T (Fig. 3c). This behavior is opposite to what is found in Sr$_3$Ru$_2$O$_7$ where $T_\mathrm{FL}$ decreases in the magnetic field in response to the field-enhanced spin fluctuations driven by the proximity to a metamagnetic critical point. In CaRuO$_3$, the increase of $T_\mathrm{FL}$ in magnetic field rather suggests a suppression of the magnetic fluctuations and the associated spin entropy. This may hint at the proximity to a zero-field QCP.

We now address the data beyond the FL regime. Above $T_\mathrm{FL}$, the resistivity crosses over from a quadratic to a milder $T^{3/2}$ dependence, as evidenced in Fig. 3a where $\rho$ plotted vs. $T^{3/2}$ appears as a straight line in a broad range up to $\sim 25$~K. The NFL contribution to the electrical resistivity in this temperature regime is large compared to the residual resistivity of our high-quality thin film (at 25~K, $\Delta\rho\sim 7\rho_0$, cf. Fig.~3a). Such behavior is characteristic for clean NFL metals~\cite{Mathur,Gegenwart99,Custers,Paglione,Nakatsuji,Pfleiderer,Mackenzie}.
Importantly, it supports the idea that the NFL behavior is intrinsic to CaRuO$_3$ and not dominated by disorder, e.g. magnetic clusters in a paramagnetic environment~\cite{Demko}. Interestingly, the exponent of 1.5 is similar as found in the NFL state of MnSi~\cite{Pfleiderer} but different to the expectation from the itinerant Hertz-Millis-Moriya theory for 3D ferromagnetic quantum criticality~\cite{vL}. Unlike the lower boundary of the $T^{3/2}$ regime, $T_\mathrm{FL}$, its upper boundary ($\sim 25$~K) is found to be field independent.

\begin{figure}
\includegraphics[width=0.95\linewidth]{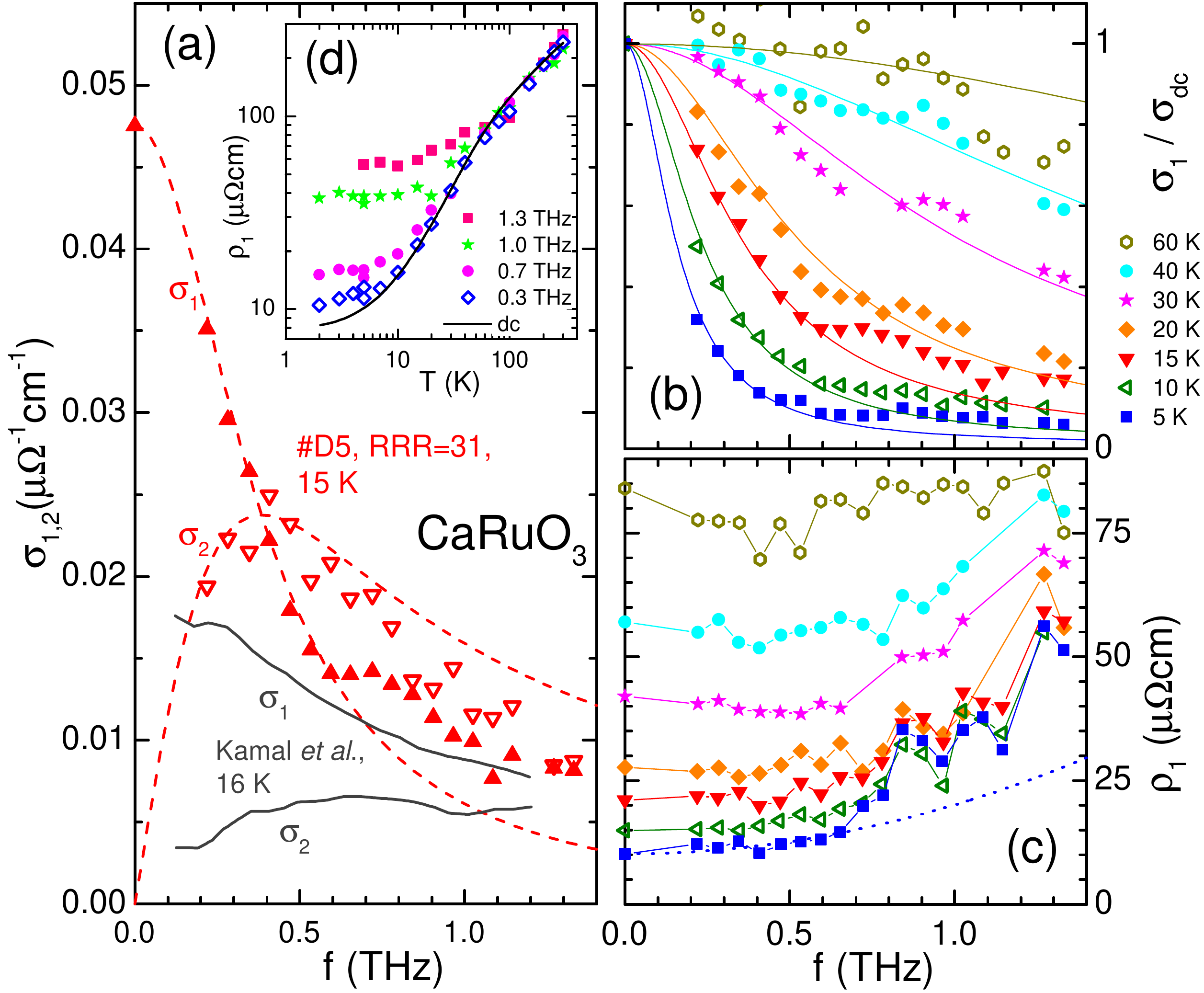}
\caption{\label{Fig_THz}(Color online) THz optical response of a 40~nm thick CaRuO$_3$ film grown on (110) NdGaO$_3$. 
(a) Real and imaginary parts of optical conductivity at 15~K (symbols). Dashed lines are a combined Drude fit to $\sigma_1$ and $\sigma_2$. For comparison, data by Kamal \textit{et al.} are plotted as full lines \cite{Kamal2006}. (b) Frequency and temperature dependence of $\sigma_1/\sigma_{\rm dc}$, compared to FL model \cite{Berthod}. (c) Frequency and (d) temperature dependence of real part $\rho_1$ of resistivity (proportional to optical scattering rate $\Gamma(\omega)$). Dashed line in (c) is a FL prediction for 5~K based on dc data below $T_{\textrm{FL}}$.
}\end{figure}

We now turn to the optical properties. We have evaluated the complex optical conductivity $\hat{\sigma}(\omega)=\sigma_1(\omega) + i \sigma_2(\omega)$ from transmission and phase shift measurements 
at frequencies $0.2-1.4$~THz and temperatures $2- 
300$~K~\cite{Pracht2013} (see also SM). 
Comparable measurements on single crystals are impossible due to reflectivity very close to unity.

The improved sample quality has striking consequences on 
the THz response. In Fig.~\ref{Fig_THz}(a) we show 
$\sigma_{1,2}$ at $T=15$~K. At low
frequencies, $\sigma_{1,2}$ are fit well by a simple Drude response $\hat{\sigma}=\sigma_{\textrm{dc}}/(1 -i \omega\tau_{\textrm{D}})$ 
with Drude relaxation time $\tau_{\textrm{D}}$. The crossing of the $\sigma_1$ and
$\sigma_2$ spectra, at $(2\pi \tau_{\textrm{D}})^{-1}= 0.4$~THz, signals the transition from the low-frequency
dissipative (transport) regime to the inductive (relaxation) regime. For our lowest temperatures, we obtain $\tau_{\textrm{D}} =$ 1~ps, in agreement with the SdH and the dc resistivity values  discussed above.
At frequencies above 0.6~THz, $\sigma_1$ clearly deviates from the Drude prediction. On the high frequency side, $\sigma_1$ and $\sigma_2$ meet again,
indicating that the scattering has increased strongly. Comparison with data from Ref.\ \cite{Kamal2006}, also shown in Fig.~\ref{Fig_THz}(a), where such dynamic
effects can be resolved only with great difficulty, highlights the role of sample quality for these studies.

In Fig.~\ref{Fig_THz}(b) we show the frequency dependence of $\sigma_1/\sigma_\mathrm{dc}$ for a set of low temperatures. Upon cooling, the electron-electron contribution to the low-frequency scattering is diminished and the low-frequency Drude roll-off becomes increasingly sharp. At lowest temperatures measured, the narrowing saturates as the scattering due to electron-electron interactions
becomes comparable to the impurity contribution.  In the NFL
temperature range ($2-25$~K), the THz data at frequencies below 0.6~THz can be modeled well by a simple Drude response with an ansatz $1/\tau_{\textrm{D}} = a
T^{3/2} + 1/\tau_0$, with $\tau_0$= 1.3~ps, whereas at higher frequencies all spectra show pronounced deviations from Drude behavior (see SM).

Such deviations were recently discussed by Berthod \textit{et al.} who have, assuming FL self-energy, derived a universal dependence of the optical conductivity in terms of a two-parameter scaling function \cite{Berthod}. In particular, a non-Drude foot in $\sigma_1$ and two crossings between $\sigma_1$ and $\sigma_2$ were discussed there, in qualitative resemblance to our data. In the low-frequency regime, Drude behavior is recovered. As shown in Fig.\ \ref{Fig_THz}(b), our overall data can be fit remarkably well with this theory, yielding a coherence scale $T_0=150$~K and an impurity scattering $Z\Gamma=0.3$~meV. However, the agreement is only in the Drude part. Substantial deviations from Drude dynamics, predicted by that theory, occur only at frequencies too large to directly account for the unusual behavior we find here~\cite{SM}.

To describe these deviations, an extended Drude analysis $\hat{\sigma} = D(\omega)/(\Gamma(\omega) -i \omega)$ can be used, where a frequency-dependent scattering rate $\Gamma(\omega)$ replaces the constant $1/\tau_{\textrm{D}}$ \cite{Scheffler2013}.
In Fig.~\ref{Fig_THz}(c) and (d) we evaluate $\Gamma(\omega)$ via $\rho_1(\omega) = \textrm{Re}(1/\hat{\sigma}(\omega)) \propto \Gamma(\omega)$. 
For 5~K and frequencies below 0.6~THz, $\rho_1$ is almost constant, but for higher frequencies there is a pronounced increase of $\rho_1$ due to electronic interactions. This abrupt increase is incompatible with the quadratic FL prediction, calculated from the dc transport $A$ prefactor below $T_\mathrm{FL}$, as indicated by the dotted line in Fig.\ \ref{Fig_THz}(c), while the data below 0.6~THz is in perfect agreement with the FL increase. A weaker power-law exponent 3/2, that describes well the temperature dependence in this regime, is even more incompatible with the data above 0.6~THz. For higher temperatures the strong frequency dependence of $\Gamma(\omega)$ is lost and $\rho_1$ basically coincides with the dc resistivity throughout our frequency range, as evident from Fig.\ \ref{Fig_THz}(d).

To conclude, we have grown high-quality thin films of CaRuO$_3$ which display for the first time for this ruthenate quantum oscillations. Electrical resistivity indicates an extended NFL regime below 25~K that crosses over to FL behavior below 1.5~K. The optical data reveal a Drude response that at frequencies below 0.6 THz can be modeled with FL concepts. Pronounced deviations from FL behavior are found above 0.6 THz where an abrupt increase of the scattering might signal the coupling to soft spin-fluctuations, a scenario that could be tested by looking at the optical properties as a function of magnetic field.

We thank H. Schuhmann and M. Seibt for their assistance using the transmission electron microscope and M. Dressel for supporting the THz study. The work was supported by the DFG through SFB 602 and FOR 960. JM acknowledges the Slovenian research agency program P1-0044.

\vspace{1 cm}
\textbf{SUPPLEMENTAL MATERIAL}

\section{Thin film synthesis}

Thin films of CaRuO$_3$ (CRO) have been prepared by utilizing a metalorganic aerosol deposition technique~\cite{MAD}. Here, a solution of commercial acetylacetonates of Ca$^{2+}$ and Ru$^{3+}$, dissolved in dimethlyformamide, is sprayed with a pneumatic nozzle with dried air onto a heated substrate~\cite{Melanie_MAD}. In the hot region close to the substrate, the acetylacetonates decompose (above 250$^\circ$C), the metal ions oxidize and the oxide film grows on the substrate. In our previous work on the system Sr$_{1-x}$Ca$_x$RuO$_3$ we have used (100) oriented SrTiO$_3$ (STO) as substrate and obtained after optimization of the growth conditions (110) oriented thin films of high quality with RRR=29 ($x=0$) and 16 ($x=1$)
~\cite{Melanie_MAD,Schneider}. Careful x-ray diffraction (XRD) analysis has revealed that these epitaxial CRO films are fully strained. While STO is well suited for films with composition between x=0 and 0.5, the larger lattice mismatch of $-2\%$ for $x=1$  prevents higher RRR values for pure CRO. Thus, we have chosen NdGaO$_3$ (NGO) as substrate for the growth of high-quality CRO thin films for which the lattice mismatch is considerably smaller ($-0.5\%$). Very similar growth parameters as reported earlier~\cite{Melanie_MAD} with a substrate temperature slightly exceeding $1000^\circ$C have enhanced the RRR values up to 35. In order to further improve the thin film quality and RRR value, we have chosen vicinal NGO substrates with a 3$^\circ$ miscut angle, which optimizes regular step bunching growth. The surface morphology studied by scanning tunneling microscopy (see Fig. 1b of main paper) has proven a step bunching growth corresponding to the steps of the NGO substrate.

\section{Structural properties}

\begin{figure}[t!]
\includegraphics[width=1\columnwidth]{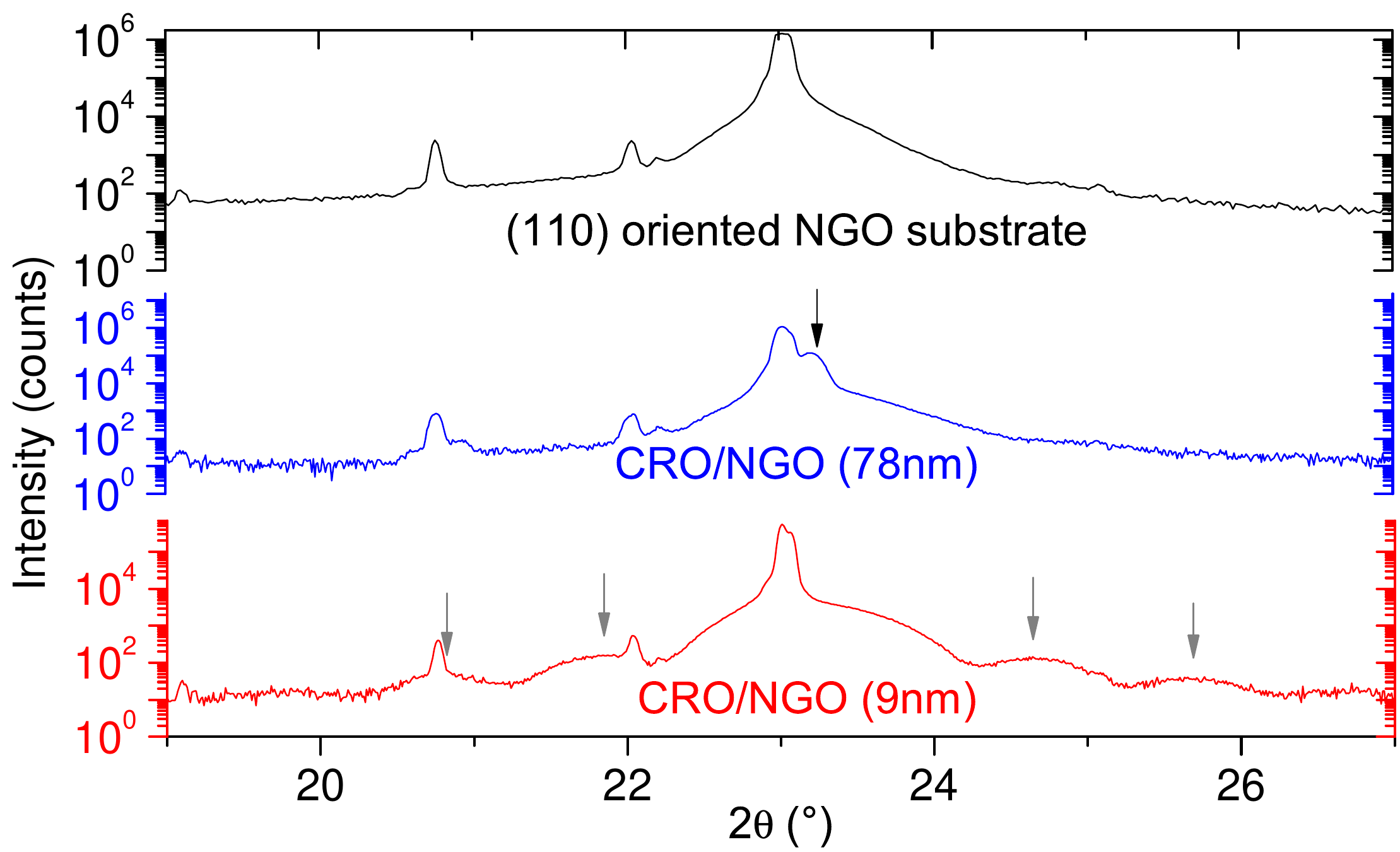}
\caption{$\Theta-2\Theta$ XRD scans around the (110) reflex of plain vicinal NdGaO$_3$ (NGO) substrate (upper panel), CaRuO$_3$ (CRO) thin film (thickness 78 nm) on vicinal NGO (middle panel) and CRO (9 nm) on vicinal NGO (lower panel). The black arrow indicates the CRO 110 reflex while the grey arrows mark Laue fringes in the very thin film.}
\end{figure}

Fig.\ 1 displays XRD scans of CRO on a vicinal NGO substrate. Here, the top panel shows the intensity spectrum of the plain NGO film while the two lower panels display CRO thin films grown on NGO. The black arrow marks the (110) reflexes from CRO in the orthorhombic Pbnm perovskite structure. No secondary phases are visible. For the thinnest film of thickness 9~nm, Laue fringes are visible which reflect a very low surface roughness. 

\begin{figure}[t!]
\includegraphics[width=1\columnwidth]{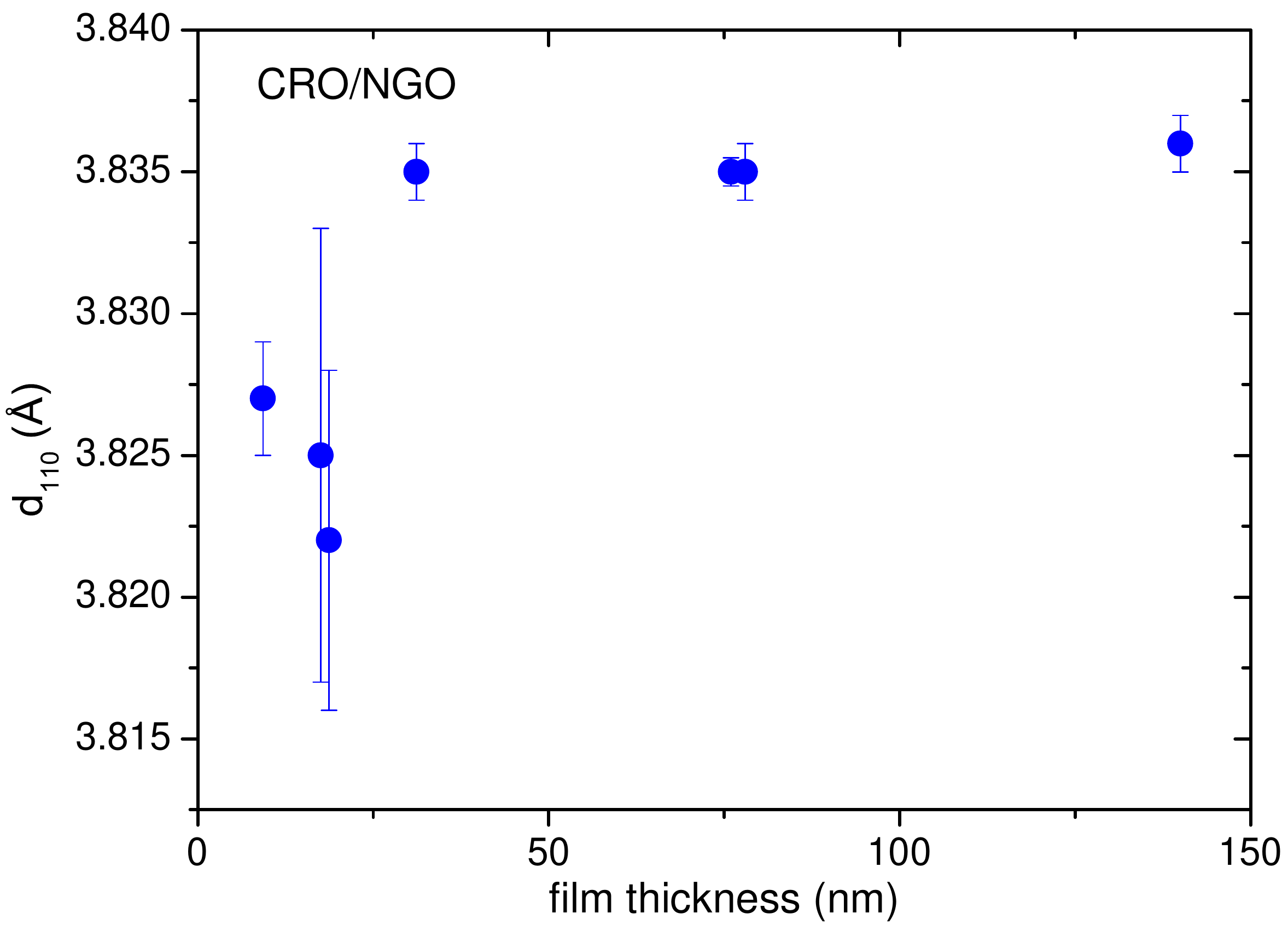}
\caption{Dependence of the (110) lattice constant for CRO thin films on vicinal NGO as function of film thickness, determined by small angle XRD.}
\end{figure}

Fig.\ 2 displays the (110) lattice constant as function of thin film thickness.  NGO exhibits a slightly bigger lattice constant compared to CRO which leads to a small in-plane tensile strain acting on the CRO thin films. This results in a decrease of the out-of-plane lattice constant, observed particularly for CRO films of low thickness (cf.\ Fig.\ 2).  A relaxation of the in-plane tensile strain caused by the negative lattice mismatch of $\sim-0.5\%$ is reached for film thicknesses larger than 20 nm. 

\begin{figure}
\includegraphics[width=0.95\linewidth]{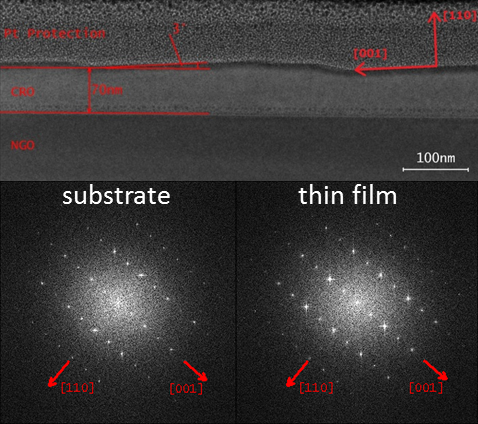}
\caption{(color online) Upper part: transmission electron microscopy (TEM) image on $\mu$m scale of the cross-section of a lamella with NdGaO$_3$ (NGO) substrate, CaRuO$_3$ (CRO) thin film and Pt protection layer (from bottom to top). The $3^\circ$ sawtooth surface of the thin film is marked by red lines. The red arrows indicate crystallographic orientations as determined by Fourier transforms of atomic-resolution TEM images of the substrate (lower left) and thin film (lower right).}
\label{Fig2}
\end{figure}

High-resolution transmission electron microscopy (TEM) measurements, displayed in Fig.\ 3, indicate homogeneous growth of CRO without any indication for secondary phases or grain boundaries. The same holds also true for studies on larger lateral dimension. The calculated fast Fourier transform of the cross-sectional TEM images indicate well ordered pattern and epitaxial growth with equal orientation of the film and substrate.

\begin{figure}[t!]
\includegraphics[width=1\columnwidth]{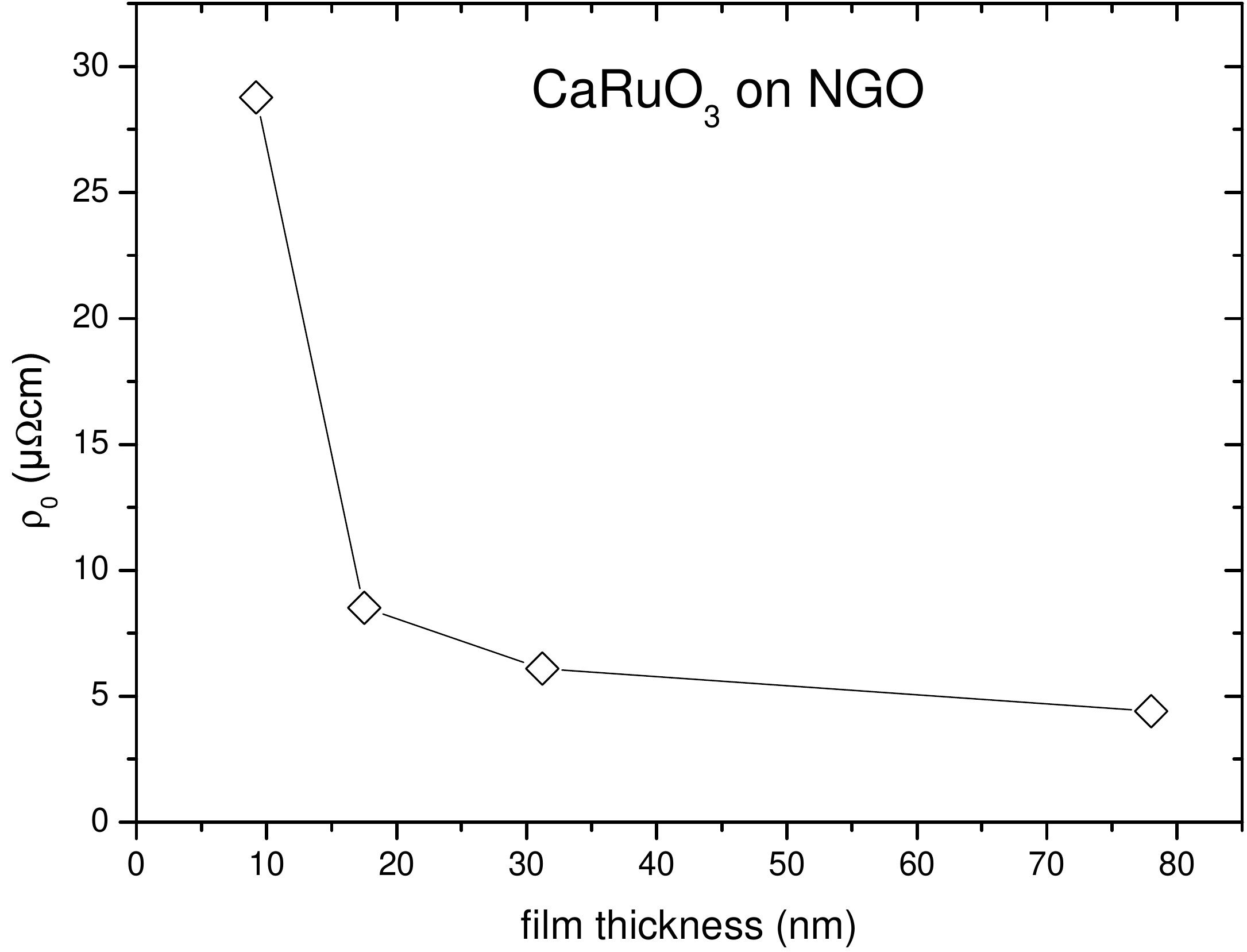}
\caption{Residual resistivity of CRO thin films on NGO substrates vs. film thickness. The room temperature resistivity amounts to $230\ \mu\Omega$cm.}
\end{figure}

As shown in Fig. 4, the residual resistivity increases with reduced film thickness. This may be attributed to surface scattering and/or the effect of in-plane tensile strain, which may act slightly inhomogeneously.

\section{Shubnikov-de Haas experiments}

For studying the Shubnikov-de Haas oscillations in the electrical resistance of CaRuO$_3$, two different thin films were micro-structured~\cite{Schneider}. The bar width of sample D12 is $50\mu m$, that of sample D41 is $100\mu m$. Dilution refrigerators were utilized for the four-probe resistivity measurements. The first set of data has been obtained on sample D10 in a superconducting laboratory magnet with maximum field of 18 Tesla, applied perpendicular to the thin film plane along the [110] direction. Subsequently, we have performed high-field measurements at the LNCMI Grenoble in an electromagnet with maximum field of 35 Tesla. There, we have measured both thin films simultaneously. Sample D41 has been mounted such that we could study the variation of the quantum oscillations upon rotating the field from the [110] towards the [001] direction. However, as the field is rotated into the thin film plane for an angle larger than $40^\circ$, the SdH oscillations are suppressed because the cyclotron orbits interfere with the sample edges.

\section{Fermi surfaces in band-theory}

We have calculated the electronic structure of CaRuO$_3$ within the density-functional theory using the local-density approximation (LDA) as implemented in the Wien2k package~\cite{wien2k}. The structural information was taken from Table I in Ref.~\cite{zayak}.  The electronic structure of CaRuO$_3$ was previously calculated and was in
detail discussed~\cite{Mazin}, however, the detailed information on its Fermi surfaces was not given. Using the SKEAF program~\cite{skeaf} we have calculated also the extremal orbits and the corresponding
frequencies of quantum oscillations and their effective masses. We visualized the surfaces and the extremal orbits using python scripts that used Mayavi libraries~\cite{mayavi}. We checked our isualizations against XCrysden~\cite{xcrysden} for consistency. To verify that the residual resistivity is compatible with the band theory we calculated the conductivity using the Boltzmann transport approach described in Ref.~\cite{boltztrap} and obtained
$\sigma_\mathrm{dc}/\tau_\mathrm{tr}=0.15\times 10^{21}$/$\Omega$m s.

\begin{figure}
\includegraphics[width=0.95\linewidth]{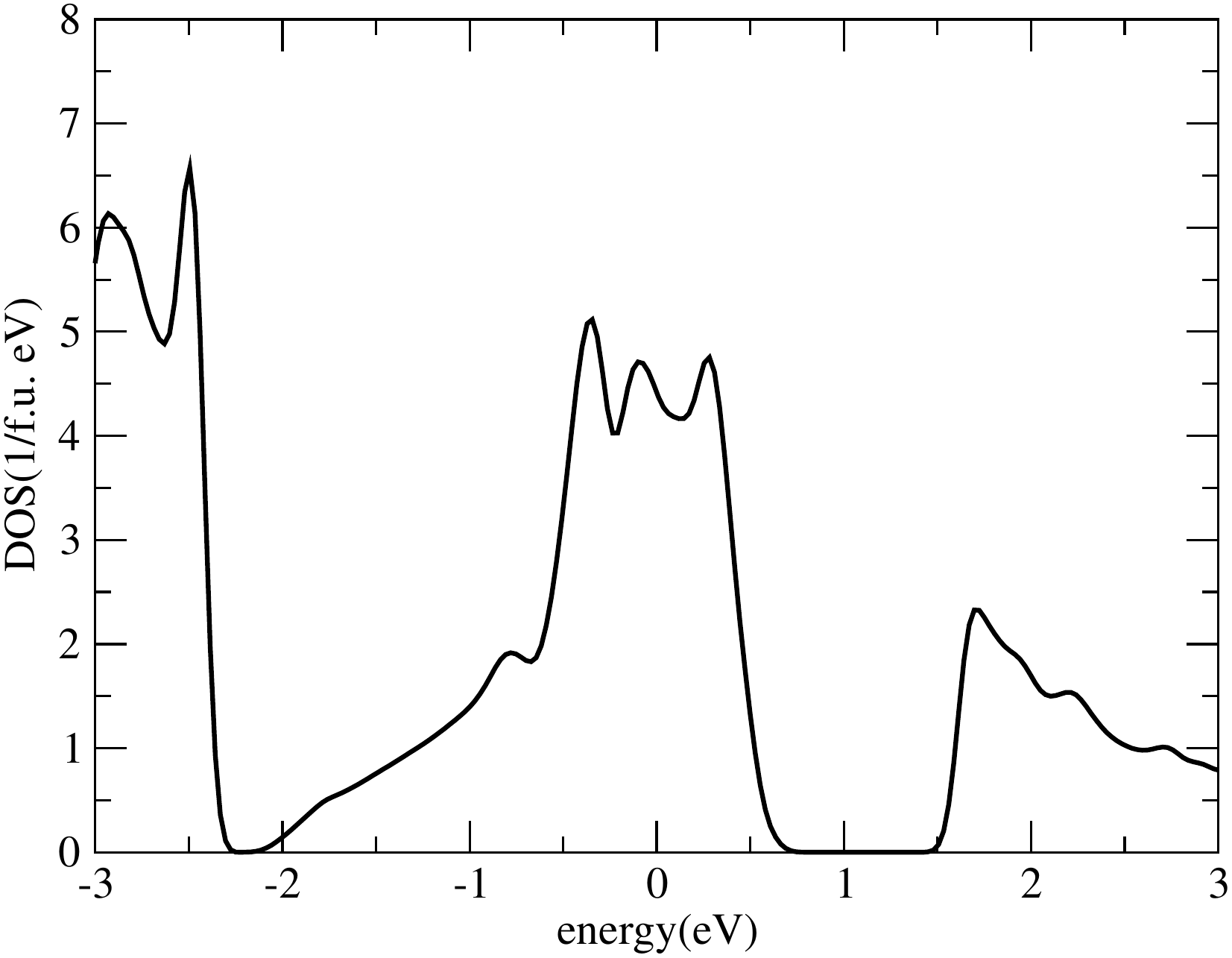}
\caption{Calculated density of states (DOS) for CaRuO$_3$. The 
energy is measured with respect to the Fermi energy.}
\label{fig:dos}
\end{figure}

\begin{figure}
\includegraphics[width=0.95\linewidth]{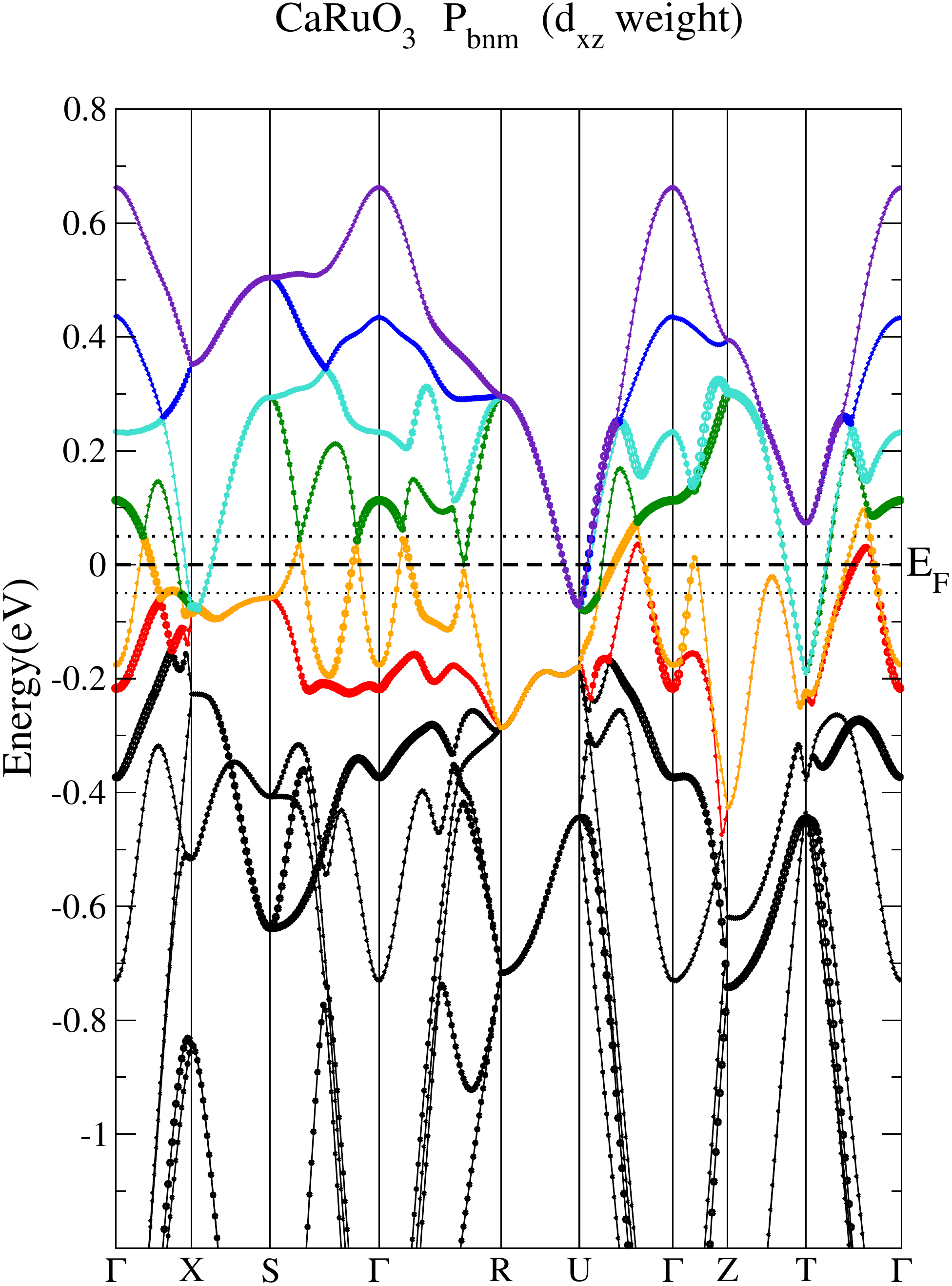}
\caption{LDA band structure close to Fermi energy in CaRuO$_3$.  The d$_{xz}$
  weight is indicated by the thickness of the symbols. 
  The points along the path are (in units of
  reciprocal unit wave-vectors) $\Gamma=(0,0,0)$,
  X=$(0,\frac{1}{2},0)$, S=$(\frac{1}{2},\frac{1}{2},0)$,
  R=$(\frac{1}{2},\frac{1}{2},\frac{1}{2})$,
  U=$(0,\frac{1}{2},\frac{1}{2})$, Z=$(0,0,\frac{1}{2})$,
  T=$(\frac{1}{2},0,\frac{1}{2})$.  Fermi level (dashed) and its
  possible spin-dependent shifts in the magnetic field of 10~T, taking Stoner
  enhancements into account (dotted) are indicated.
\label{fig:bands}}
\end{figure}

The CaRuO$_3$ DOS is shown on Fig.~\ref{fig:dos}. The DOS at the Fermi energy is 4.41/f.u. eV, which corresponds to a linear coefficient in specific heat $\gamma=10.4$mJ/molK$^2$. The low energy (``t$_{2g}$'')
manifold is gapped from the higher energy (``e$_g$'') manifold due to orthorhombic splittings.  CaRuO$_3$ electronic dispersions are shown on Fig.~\ref{fig:bands}. As there are 4 formula units in the orthorhombic P$_\textrm{bnm}$ unit cell, there are 12 ``t$_{2g}$'' bands that have also some e$_g$ character due to orthorhombic distortions. In the algorithm, the bands are numbered according to their energy at each
k-point separately, thus the band-character is not neccesarily
followed properly along the k-path. All the crossings between bands are mistaken for anticrossings, as can be seen by following the orbital character that is indicated by the symbol size on the plot. This has no consequences for the identification of Fermi surfaces.

\begin{figure}
\begin{overpic}[trim=0 0 0 0,clip=true,width=0.49\linewidth]{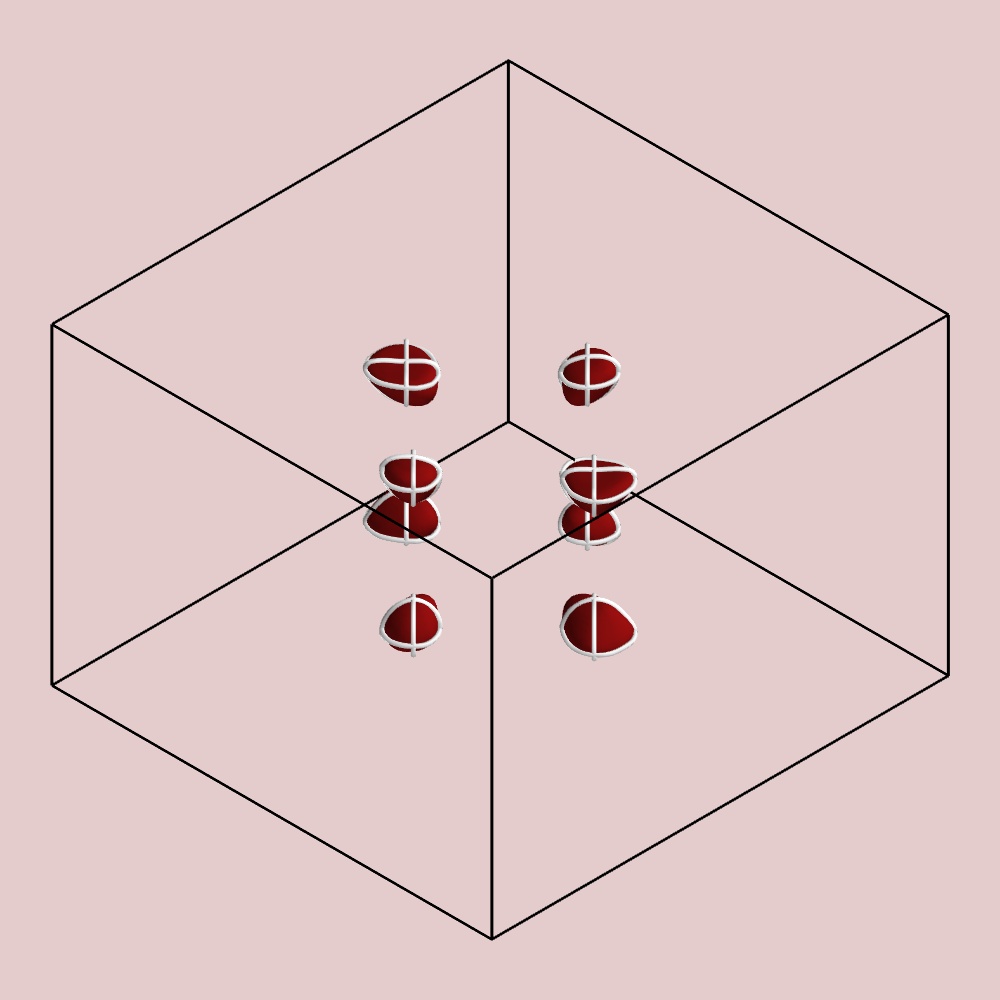}
\put(65,30){$\alpha$}  \put(87,6){(a)}
\end{overpic}
\begin{overpic}[trim=60 60 130 130,clip=true,
    width=0.49\linewidth]{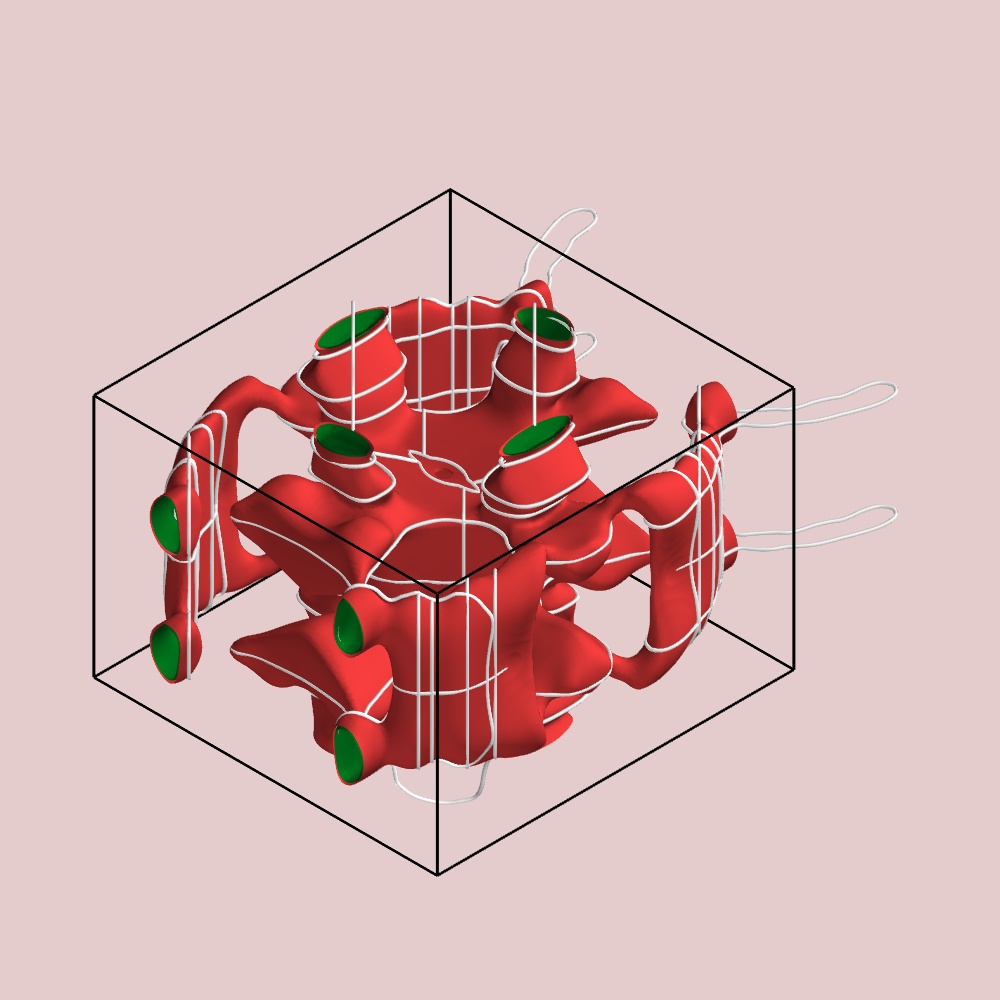} \put(70,25){$\zeta$}
  \put(87,6){(b)}
\end{overpic} \\
\begin{overpic}[trim=90 70 130
    140,clip=true,width=0.49\linewidth]{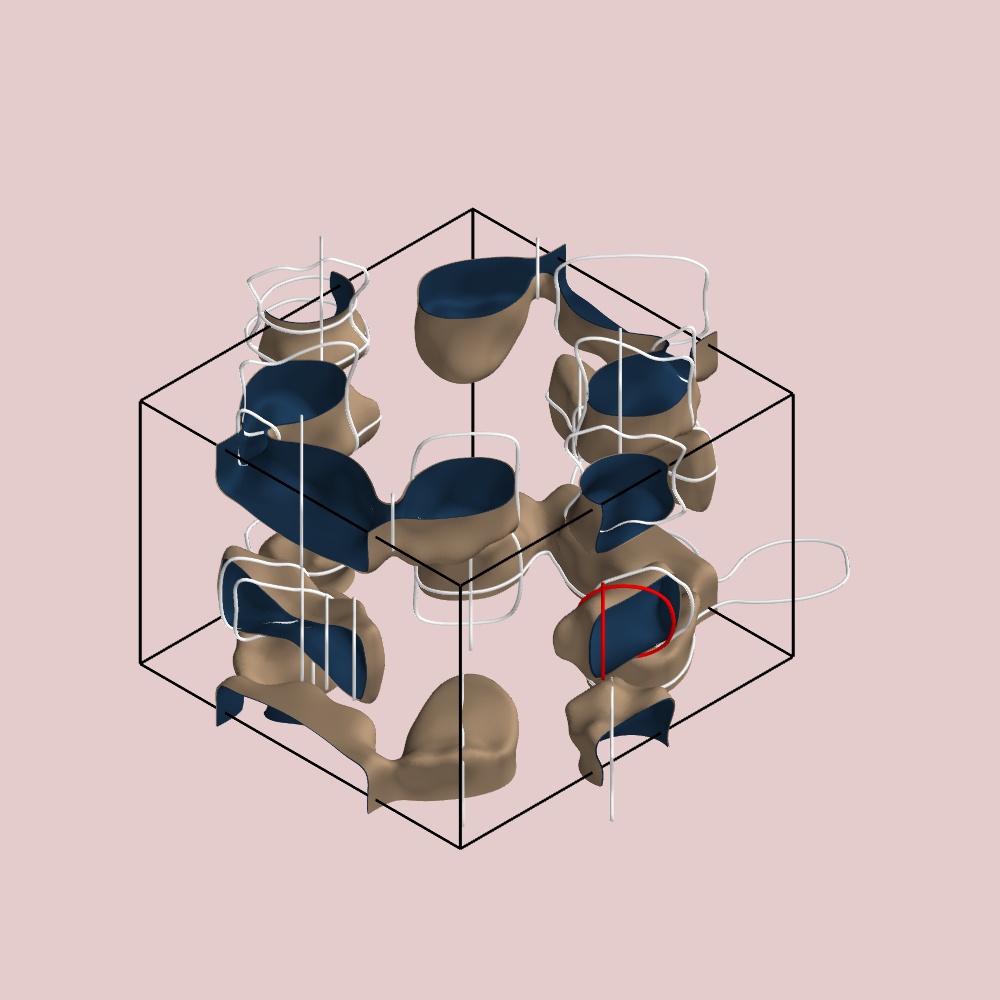}\put(70,16){$\beta_4$}
  \put(36,31){$\gamma_2$} \put(54,32){$\delta_2$}
  \put(23,18){$\epsilon_2$}  
\put(87,6){(c)}
\end{overpic}
\begin{overpic}[trim=50 30 50 60,clip=true,width=0.49\linewidth]{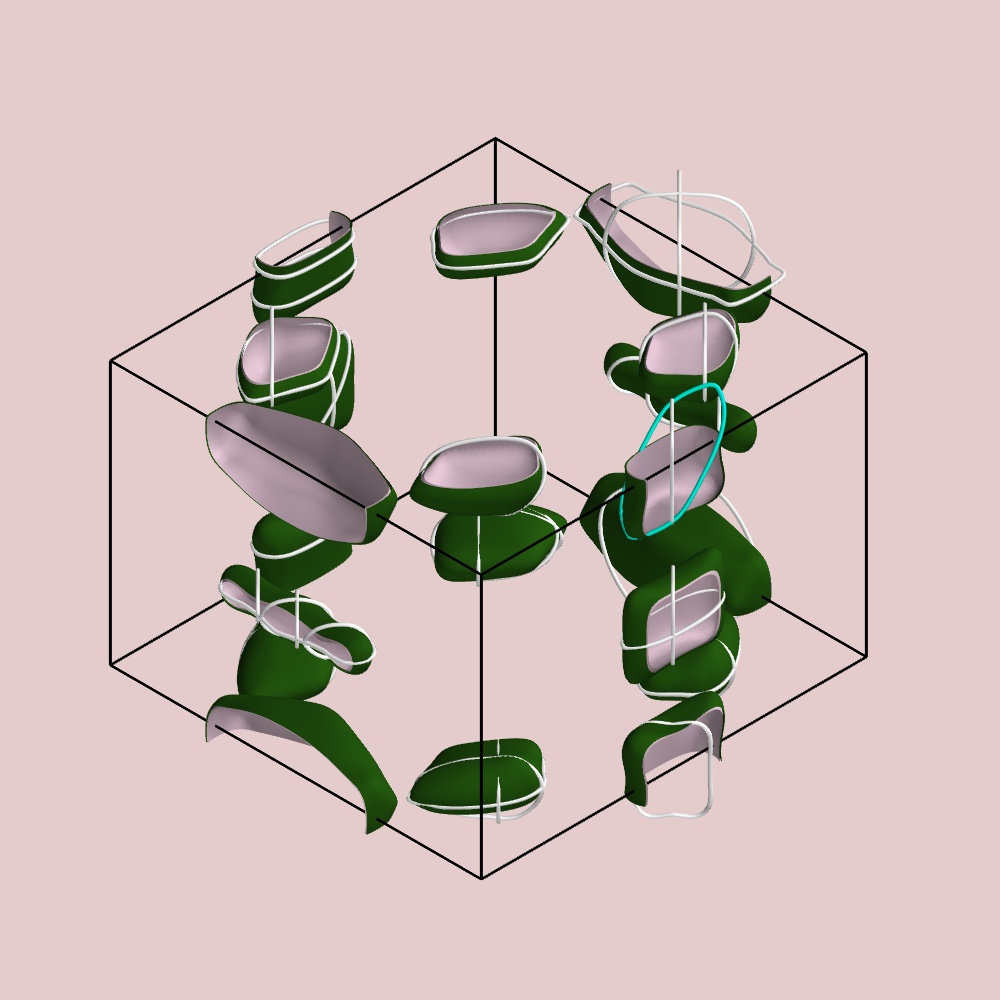}\put(73,18){$\beta_3$} \put(36,33){$\gamma_1$}  \put(57,32){$\delta_1$}  \put(52,15){$\epsilon'_1$}  \put(24,18){$\epsilon_1$}  \put(87,6){(d)}
\end{overpic} \\
\begin{overpic}[trim=30 0 30 40,clip=true,width=0.49\linewidth]{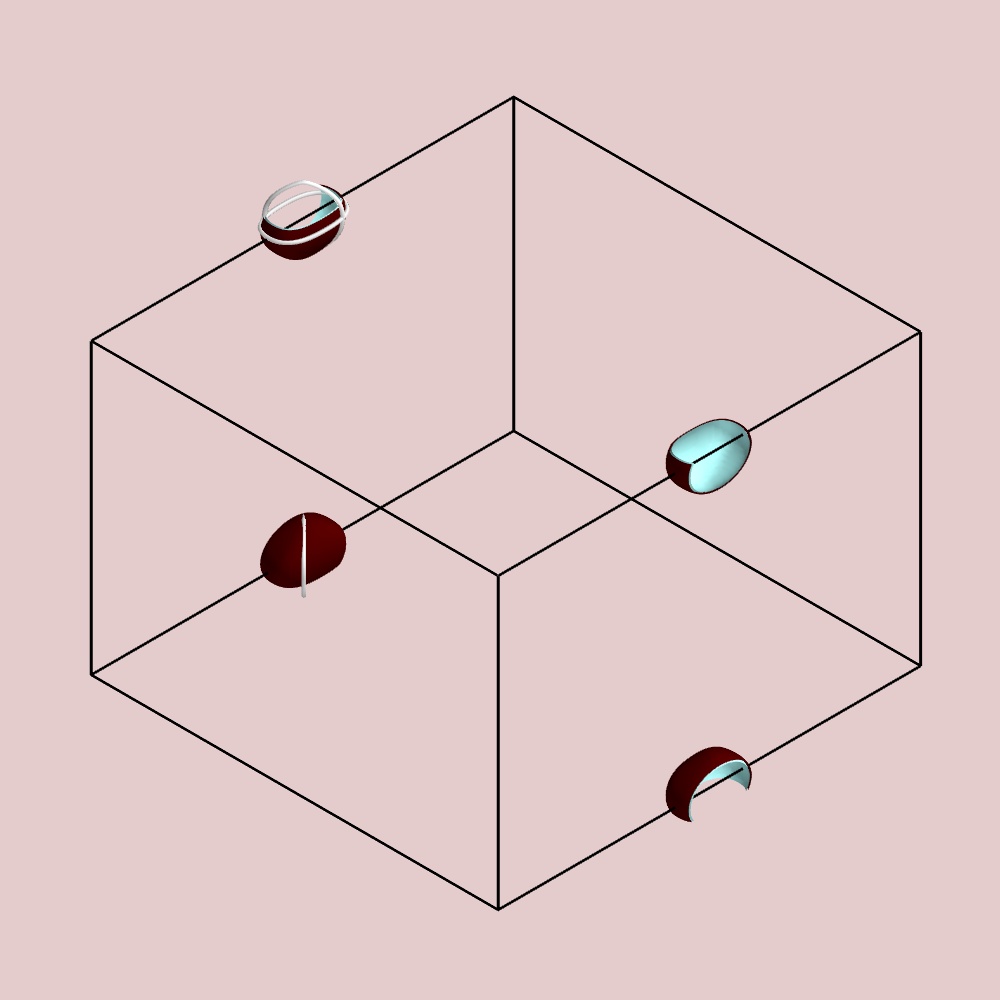}\put(73,15){$\beta_2$} \put(87,6){(e)}
\end{overpic}
\begin{overpic}[trim=30 0 30
    40,clip=true,width=0.49\linewidth]{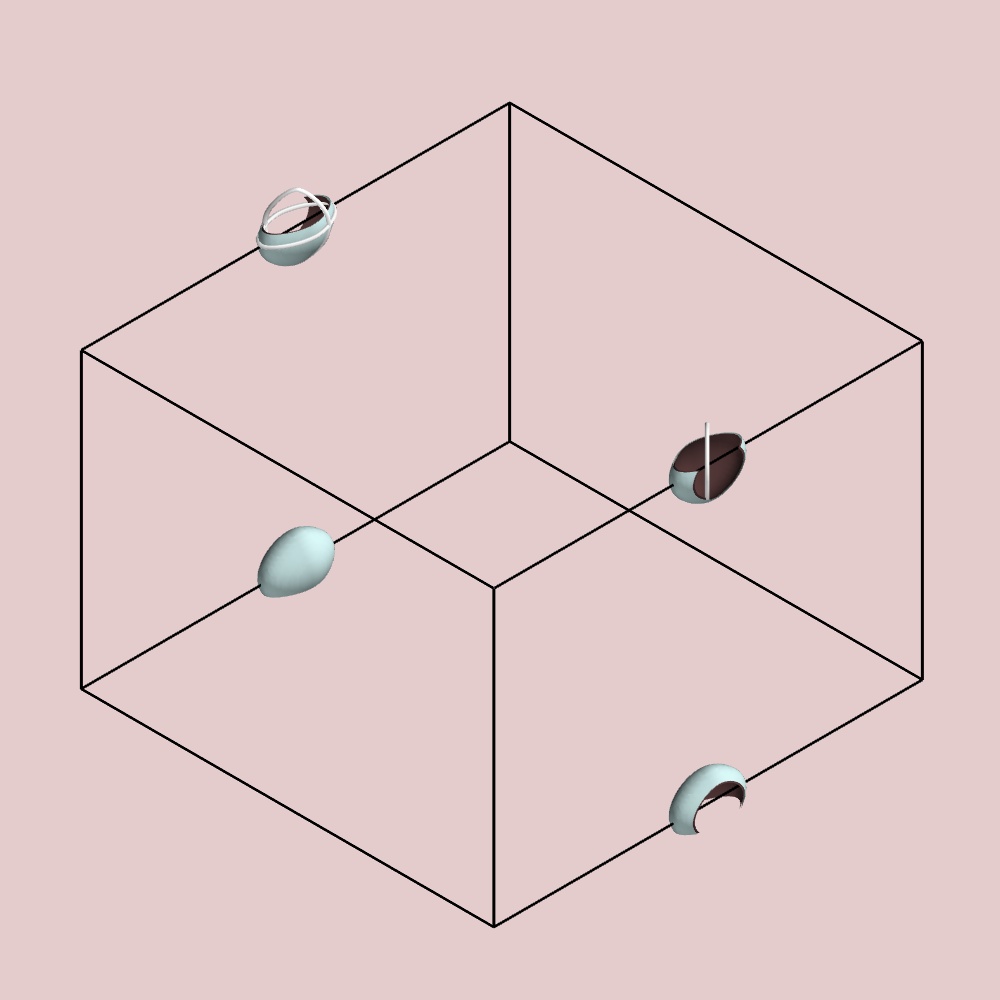}\put(74,14){$\beta_1$} \put(87,6){(f)} \put(48.5,92){$\scriptstyle \bullet \, \,(-\frac{1}{2},-\frac{1}{2},\frac{1}{2})$} \put(4,31.3){$\scriptstyle \bullet \, \,(\frac{1}{2},-\frac{1}{2},-\frac{1}{2})$}
\put(47,6.5){$\scriptstyle \bullet \, \,(\frac{1}{2},\frac{1}{2},-\frac{1}{2})$}
\end{overpic}
\caption{Fermi surfaces in CaRuO$_3$. The extremal orbits for magnetic
field in 110, 1-10, and 001 directions are also shown. In (c) the
experimentally observed orbit is indicated in red. In (d), the
experimentally observed orbit in the tilted magnetic field is
indicated in cyan.}
\label{fig:fermi}
\end{figure}

Several bands cross the Fermi energy and several Fermi surface sheets
are found.  The Fermi surfaces of CaRuO$_3$ are presented on
 Fig.~\ref{fig:fermi}. As the naming convention, we use Greek letters
 with Latin indices to identify different Fermi surface sheets where
 $\alpha$ is used for the sheet that encloses the smallest and $\zeta$
 the largest volume. Several sheets of the Fermi surface fully enclose
 another sheet of the Fermi surface of a smaller volume. This is the case
 for instance for sheets $\beta$. In such cases we label all the
 corresponding sheets with the same Greek letter, but use Latin
 number indices to distinguish the sheets among themselves, e.g. $\beta_1$,
 $\beta_2$, $\beta_3$, $\beta_4$ for the innnermost until the
 outermost sheet, respectively.

\begin{figure}
  \begin{overpic}[trim=100 50 100 50,clip=true,width=0.99\linewidth]{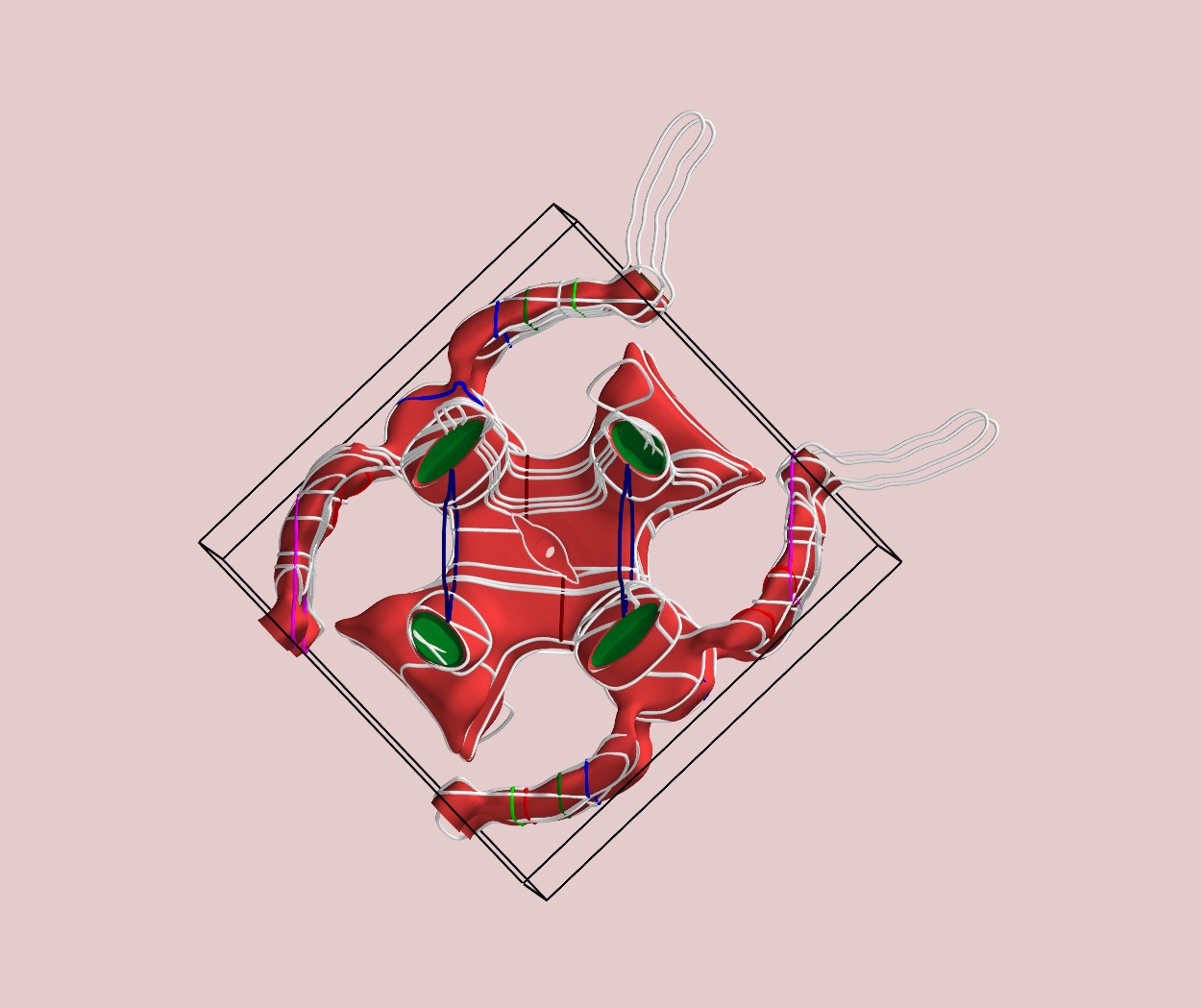}
\end{overpic}
\caption{$\zeta$ Fermi surface sheet looked at
  from 001 direction. Note the hole in the center.}
\label{fig:fermi2}
\end{figure}

On Fig.~\ref{fig:fermi}(a) one can see a small hole pocket, that one
encounters in the direction $\Gamma \rightarrow$U, and on
Fig.~\ref{fig:fermi}(e,f) two small electron pockets $\beta_1$ and
$\beta_2$, that are at the edge of the Brillouin zone, touching the
surface perpendicular to 100 direction. Figs.~\ref{fig:fermi}(c,d)
show more complex surfaces, where each sheet displayed in (d) has its
companion with a larger volume displayed in (c). The sheets
$\epsilon_1$ and $\epsilon'_1$ that appear to be separate on (d),
merge in their larger volume realization $\epsilon_2$ shown on (c),
which is also why we used $\epsilon_1'$, to label one of the two
$\epsilon$-sheets presented in (c). Sheet $\gamma_1$ has a peculiar pea-pod
(containing three 'peas') shape and is enclosed by a more irregular
$\gamma_2$ sheet. Center of the face 010 is enclosed in two
flat-pebble-like sheets, $\delta_{1,2}$. Fig~\ref{fig:fermi}(b)
displays $\zeta$ sheet with a complicated shape and
topology, see also Fig.~\ref{fig:fermi2} for a view from the top. It
can vagely be described as consisting of a central Greek cross shape with
additional wings close to the Brillouin zone boundaries
perpendicular to the 010 direction. The central cross furthermore
contains a spaceous cave around the $\Gamma$ point in its interior, with
small ``windows'' that are visible when looked at from the 001 direction.

Some of the extremal orbits for a few orientations of magnetic field are visualized with tubes on Fig.~\ref{fig:fermi}. Two orbits whose oscillation frequencies are consistent with the frequencies that we were able to extract from the experimental signal with confidence are indicated in color: red orbits for the field in (degenerate) 110 and 1-10 directions (c) and cyan for the tilted magnetic field for 35
degrees towards the 001 direction (d).  The quantum-oscillation
frequencies, the effective masses, the hole/electron character of the
orbit and the corresponding sheet of the Fermi surface are presented
in Table~\ref{table_1} for magnetic field pointing in the 110
direction. We refer to different extremal orbits using the name of the
sheet, with additional superscript Roman number (with small numbers
for short orbits).  

\begin{table*}
\begin{center}
\begin{tabular}{c c c c c  c}
name  & sheet & ``band'' index & frequency[T] & mass [el.mass]  & comment
\\
\hline
$\alpha^1$ & $\alpha$ & 87 & 86 & 0.32 & hole pocket \\
$\alpha^2$ & $\alpha$ & 87 & 96 & 0.36 & same hole pocket, from different side \\
\hline
$\zeta_1$ & $\zeta$ & 88 &  300 & 1.44 & hole-like (wing,
next-most-far from body) \\
$\zeta_2$ & $\zeta$ & 88 &  335 & 1.18 & hole-like (wing, most-far
from body) \\
$\zeta_3$ & $\zeta$ & 88 &  338 & 2.41 & hole-like (wing, most-close
to body) \\
$\zeta_4$ & $\zeta$ & 88 &  396 & 1.48 & hole-like (goes through
``stomach'') \\
$\zeta_5$ & $\zeta$ & 88 &  434 & 1.17 & hole-like (wing, next
most-close to body) \\
$\zeta_6$ & $\zeta$ & 88 &  1089 & 1.45 & electron-like (follows necks
that penetrate out of Brillouin zone) \\
$\zeta_7$ & $\zeta$ & 88 &  1512 & 4.72 & hole-like (goes around the
wing in the long direction)\\
\hline 
$\delta_2^2$ & $\delta_2$ & 89 & 483  & 1.03 & electron pocket (flat-pebble) \\
$\delta_2^1$ & $\delta_2$ & 89 & 461  & 0.83 & electron pocket (flat-pebble) \\
$\gamma_2^3$ & $\gamma_2$ & 89 & 438 & 0.85 & electron pocket
(pea-pod: side) \\
$\gamma_2^2$ & $\gamma_2$ & 89 & 389 & 1.10 & electron pocket
(pea-pod: center) \\
$\gamma_2^1$ & $\gamma_2$ & 89 & 354 & 0.97 & electron pocket
(pea-pod: side) \\
$\epsilon_2^4$ & $\epsilon_2$ & 89 & 1019 & 1.20 & electron pocket
(neck part, different side)  \\
$\epsilon_2^3$ & $\epsilon_2$ & 89 & 799 & 1.40 & electron pocket
(side part, different side)  \\
$\epsilon_2^2$ & $\epsilon_2$ & 89 & 785 & 1.06 & electron pocket
(side part)  \\
$\epsilon_2^1$ & $\epsilon_2$ & 89 & 133 & 1.03 & electron-like
triangular-shaped neck   \\
$\beta_4$  & $\beta_4$ & 89  & 773 & 1.40 & electron pocket  \\
\hline
$\epsilon_1$ & $\epsilon_1$ & 90 & 780 & 0.75 & electron pocket  \\
$\epsilon'_1$ & $\epsilon'_1$ & 90 & 254 & 0.43 & electron pocket  \\
$\epsilon'^2_1$ & $\epsilon'_1$ & 90 & 362 & 0.76 & electron pocket  \\
$\delta_1$ & $\delta_1$ & 90 & 263 & 0.63 & electron pocket (flat-pebble) \\
$\gamma_1^2$ & $\gamma_1$ & 90 & 124 & 0.83 & electron pocket (pea-pod: center) \\
$\gamma_1^1$ & $\gamma_1$ & 90 &  104 & 0.76 & electron pocket
(pea-pod: side) \\
$\beta_3$  & $\beta_3$ & 90  & 409 & 0.86 & electron pocket (inside
$\beta_4$) \\
\hline
$\beta_2$  & $\beta_2$ & 91 & 210 & 0.55 & electron pocket (inside $\beta_3$) \\
$\beta_1$  & $\beta_1$ & 92 & 163 & 0.38 & electron pocket (inside $\beta_2$) \\
\end{tabular}
\end{center}
\caption{The data about SdH extremal orbits when the magnetic field is in the 110 direction. \label{table_1}}
\end{table*}

\begin{figure}
\includegraphics[width=\linewidth]{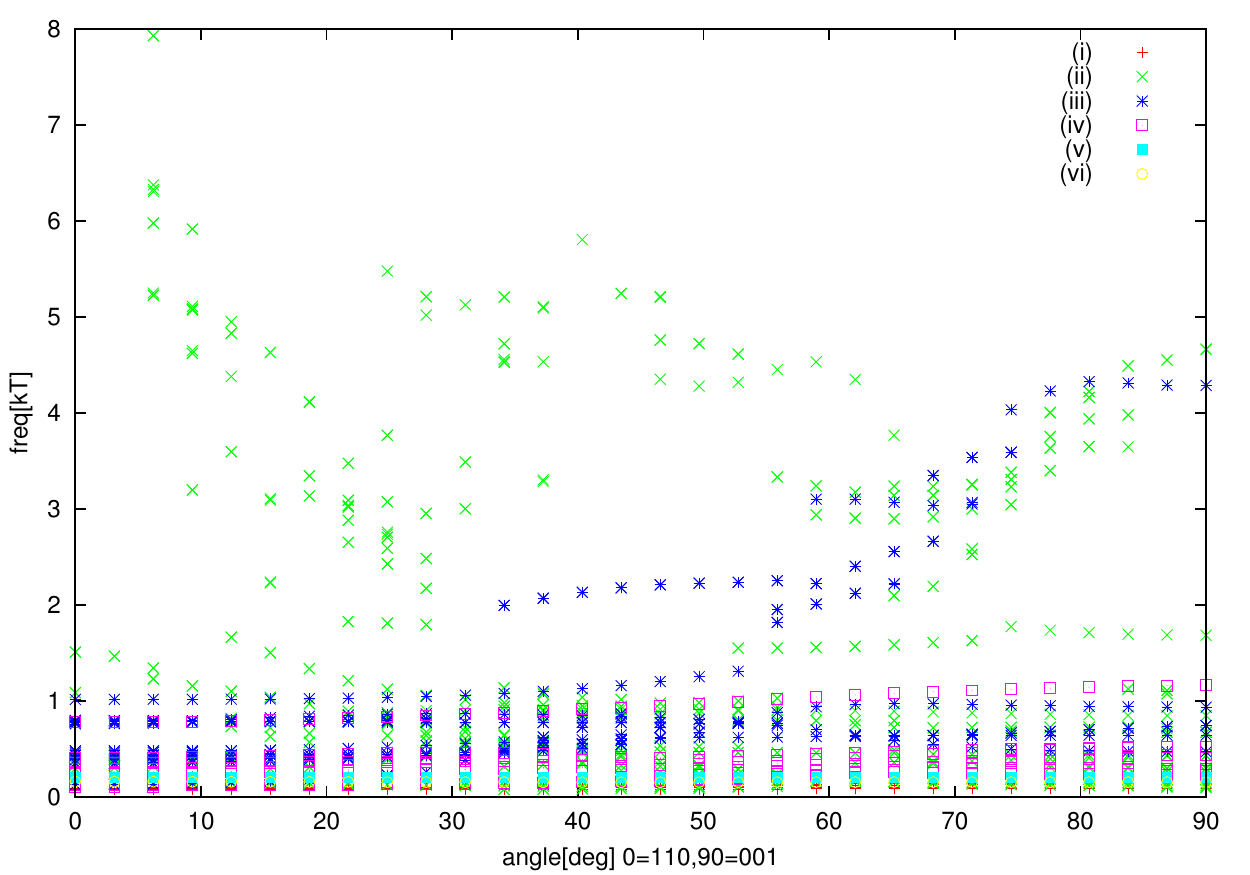} \\
\includegraphics[width=\linewidth]{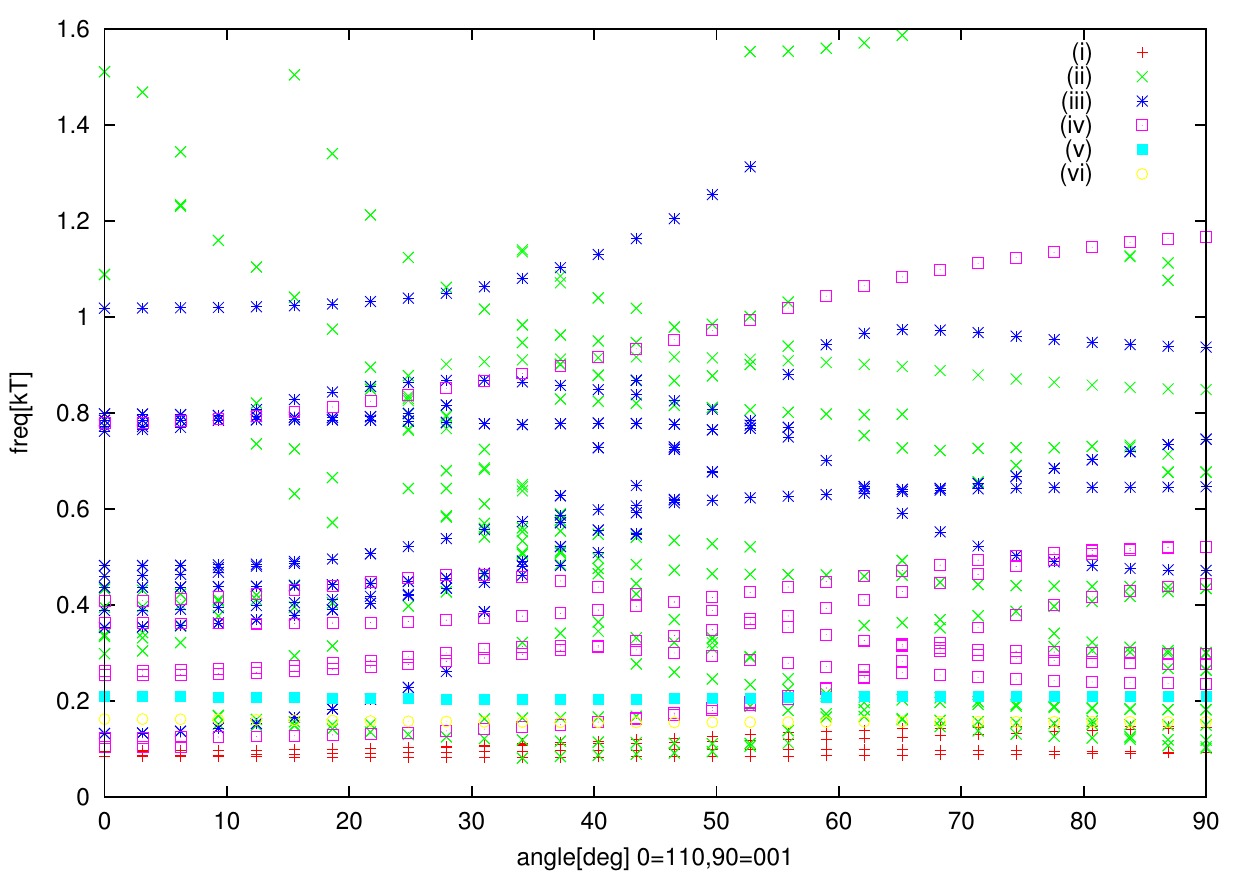} 
\caption{The angular dependence of SdH frequencies as the magnetic
  field is turned from the 110 to 001 direction. Broad (top) and
  narrower frequency range (bottom) are shown.  Different symbols
  are used for extremal orbits for the Fermi surface sheets presented
  in Fig.~\ref{fig:fermi}(a-f), for (i-vi), respectively.}
\label{fig:angular}
\end{figure}

On Fig.~\ref{fig:angular} the angular dependence of the SdH
frequencies is shown. The window from about 0.1 to about 1kT is quite
densely covered. The smallest pockets have frequencies about 80T.

\subsection{Extremal orbits at around 35 degrees inclination} 
The Fourier transformed SdH data at 35 degrees inclination from 110 to
001 reveals two well identifiable peaks: one with frequency
541$\pm$20T and mass $4.9 \pm 0.3$$m_e$ and one with frequency
472$\pm$20T and mass $4.3 \pm 0.4$$m_e$.  We associate the origin of
the signal to the 460 T extremal orbit on sheet $\beta_3$ with mass
1$m_e$. Other extremal orbits have frequencies within experimental
errorbars, but not only their angular dependence is incompatible with
our observations but also their masses are too high (1.3$m_e$ for a
460T orbit associated with $\gamma_2$ sheet which is the lightest
among them).

\section{TH\MakeLowercase{z} study}

\subsection{THz experiment and determination of optical conductivity}
We have studied the THz properties of a 40~nm thick CRO film grown on a non-vicinal (110) NGO substrate. The residual resistivity ratio of this film is 31. At this stage, we cannot employ thicker films with even higher RRR for the THz experiments because the transmitted signal would be too small to obtain sufficient signal-to-noise; the transmission of the present sample is of order 0.001 at low temperatures and frequencies. This small transmission also sets the low-frequency limit of the spectral range that we can employ reliably for this study: at low frequencies, a small fraction of the THz radiation reaches the detector without interacting with the sample. This \lq parasitic radiation\rq, which is not understood well, is probably caused by diffraction at e.g.\ apertures of cryostat windows. Its strength increases with decreasing frequency, and in our case of a low-transmission sample we can rule out significant errors due to the parasitic radiation only for frequencies above 0.2~THz.

We performed transmission and phase measurements using a set of backward-wave-oscillators (BWOs) as tunable monochromatic and coherent THz sources. Phase measurements were performed with a Mach-Zehnder interferometer. To reach cryogenic temperatures, we employed two different home-built cryostats, one of them for temperatures down to 2~K, the other one, with tilted outer windows to reduce undesired standing waves, for temperatures down to 5~K \cite{Pracht2013}.
Both transmission and phase spectra display pronounced Fabry-Perot oscillations due to standing waves in the dielectric substrate, and these can be described by formulas based on Fresnel equations \cite{Dressel2002}. Individually modeling each of these oscillations with two complex optical constants (e.g.\ $\sigma_1$ and $\sigma_2$), using values for the dielectric properties of the substrate independently obtained on a plain NGO reference, we determine very precise values for the optical properties of the CRO film.

The NGO substrate with a thickness of 0.5~mm, a dielectric constant around 22, and negligible dielectric loss at low temperature is well suited for these THz transmission measurements. This is in contrast to STO substrates, which often dominate over the influence of a metallic film and furthermore are very difficult to model \cite{Geiger2012}. (STO has a very high and strongly temperature- and frequency-dependent dielectric constant combined with large dielectric losses.)

\subsection{Optical spectra at temperatures up to 300~K}
\begin{figure}[tb]
\includegraphics[width=0.95\linewidth]{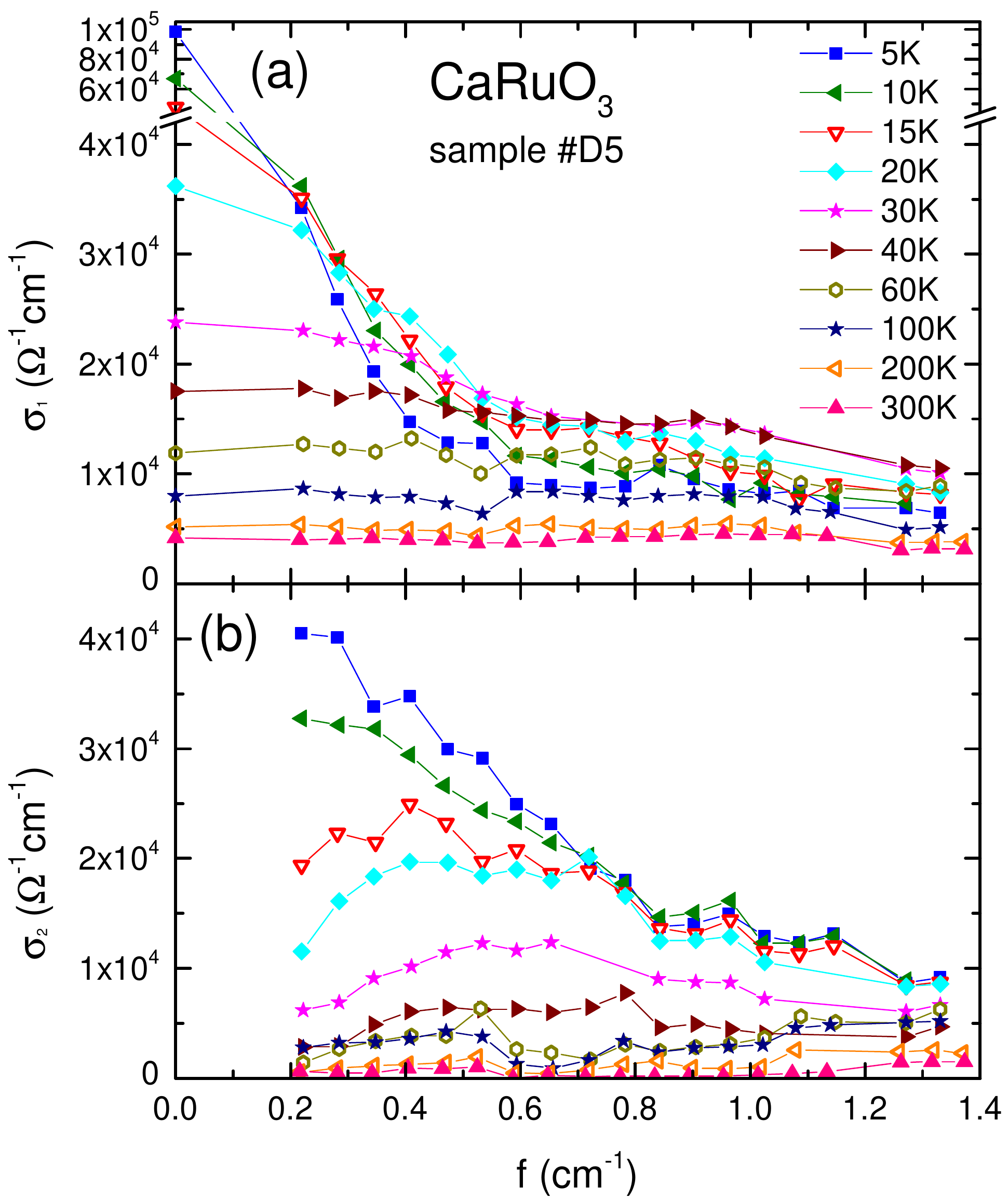}
\caption{\label{Fig_THzHighT} (a) Real part $\sigma_1$ and (b) imaginary part $\sigma_2$ of optical conductivity for our sample \#D5 at temperatures between 5~K and 300~K.}
\end{figure}

Fig.\ \ref{Fig_THzHighT} shows conductivity data for a set of
temperatures from 5~K up to 300~K. The evolution of the Drude-type
roll-off in the real part $\sigma_1$ can be tracked smoothly, see
Fig.\ \ref{Fig_THzHighT}(a): at the highest temperatures, the
scattering rate is higher than our spectral range, and as a result the
$\sigma_1$ spectra are basically flat and coincide with the dc
conductivity. Upon cooling, the scattering rate decreases and a Drude
roll-off in $\sigma_1$ develops. For temperatures lower than 40~K, the
scattering rate is clearly visible in our frequency range. In the case
of a simple temperature-dependent Drude response, the relaxation rate
as a function of temperature can be evaluated by determining the
temperature with the highest $\sigma_1$ at a given frequency. Also in
this aspect we find a smooth evolution (0.6~THz $\approx$ 30~K,
0.45~THz $\approx$ 20~K, 0.35~THz $\approx$ 15~K, 0.2~THz $\approx$
10~K) within our data. For temperatures lower than 10~K, the
scattering rate has decreased to a frequency lower than what we can
access with the present THz experiment. In the imaginary part
$\sigma_2$, shown in Fig.\ \ref{Fig_THzHighT}(b), we see the
corresponding behavior: With decreasing temperature, $\sigma_2$
continuously increases for all temperatures. Starting around 40~K, we
can identify a maximum in the $\sigma_2$ spectra which moves toward
lower frequencies upon further cooling. In the simple Drude picture,
this maximum indicates the relaxation rate, and we find consistent
frequencies compared to the behavior in $\sigma_1$. For temperatures
below 15~K, we cannot identify the position of this maximum any more
because it has moved to frequencies lower than we can reliably address in this
study.

\subsection{Optical spectra at temperatures below 5~K}
\begin{figure}[tb]
\includegraphics[width=0.95\linewidth]{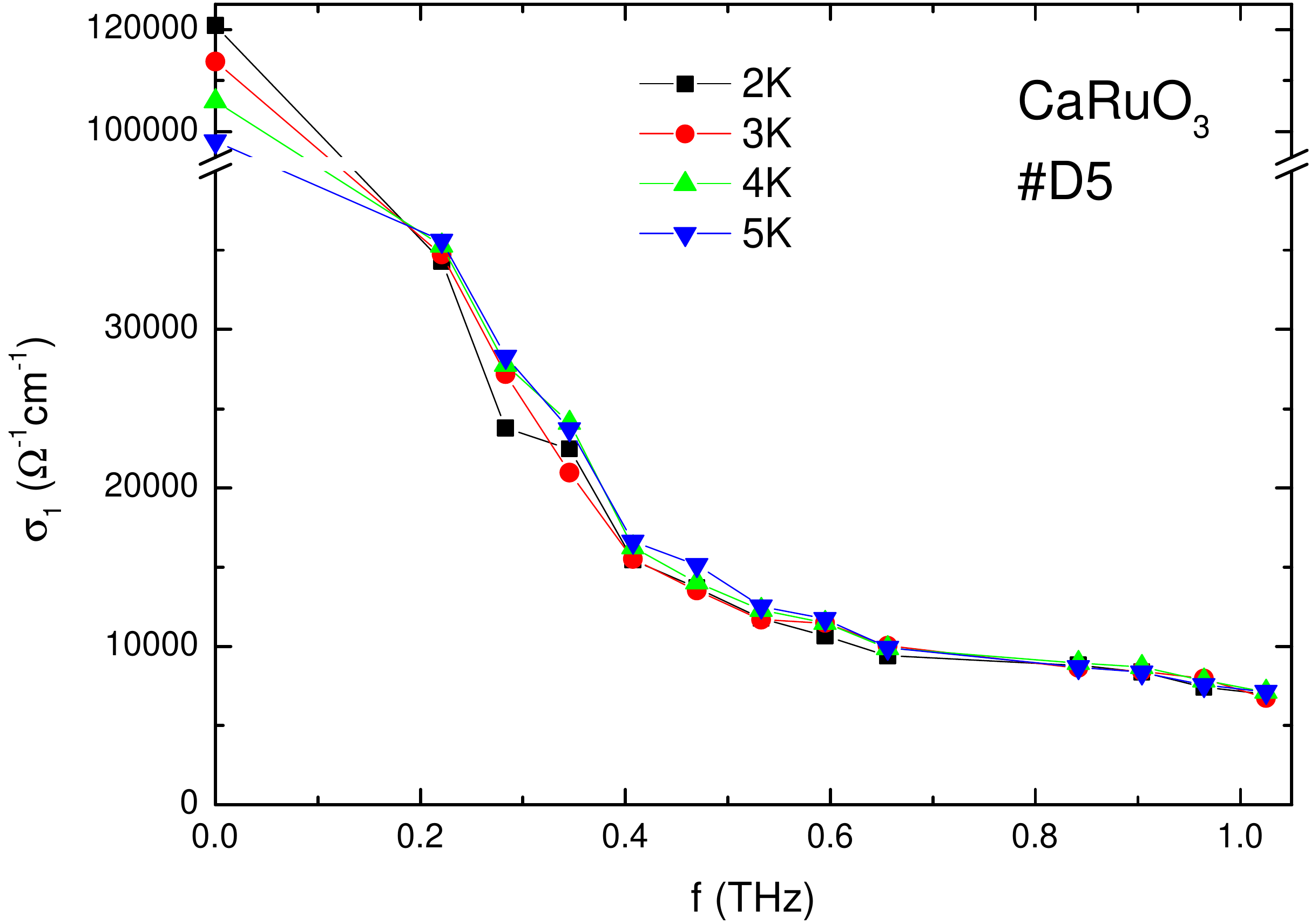}
\caption{\label{Fig_THzLowT}Real part $\sigma_1$ of optical conductivity for our sample \#D5 at temperatures between 2~K and 5~K.}
\end{figure}

The THz data for temperatures between 2~K and 5~K were performed in a separate experimental run using a home-built low-temperature optical cryostat \cite{Pracht2013}. The conductivity spectra are shown in Fig.\ \ref{Fig_THzLowT}. The trend observed at higher temperatures smoothly continues: the scattering rate reduces further, but since it is lower that our lowest accessible frequency, we observe this only by the further narrowing (toward zero frequency) of the roll-off in the $\sigma_1$ spectrum.

\subsection{Comparison with theory for a Fermi liquid}
\begin{figure}[tb]
\includegraphics[width=0.95\linewidth]{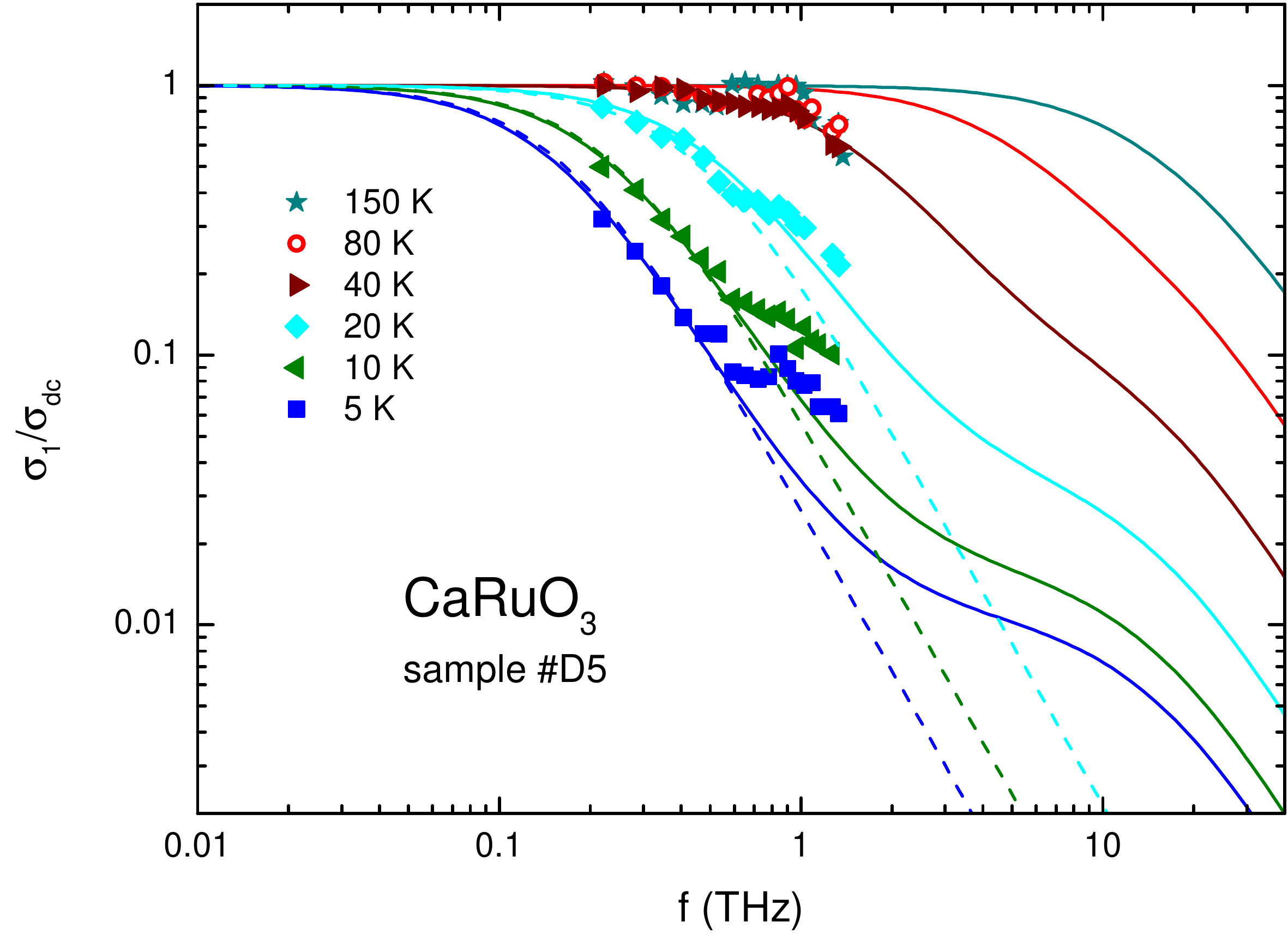}
\caption{\label{Fig_THzFLscaling}THz data for a few temperatures compared to theoretical predictions. The full lines follow a FL description \cite{Berthod}, where a non-Drude plateau is predicted around 4~THz for the 5~K curve. The dashed lines are simple Drude curves for frequency-independent scattering rate $1/\tau_{\textrm{D}}$ with NFL temperature dependence $1/\tau_{\textrm{D}} = a T^{3/2} + 1/\tau_0$.}
\end{figure}

We have compared our THz data to the theoretically expected curves for a Fermi liquid as recently discussed by Berthod \textit{et al.} \cite{Berthod}. This theory describes the optical response of a local Fermi liquid. It assumes a Fermi-liquid self energy $\mathrm\Sigma= (1-1/Z)\hbar \omega - i (\hbar^2\omega^2+\pi^2 k_B^2 T^2)/(Z \pi k_B T_0)$
 that does not depend on the wave-vector (i.e.\ is local). The vertex corrections to optical conductivity are assumed to vanish. $Z$ is the quasiparticle residue and $T_0$ is the coherence scale.

Based on these assumptions, a scaling form for the optical conductivity has been derived. The optical conductivity reads 
\begin{equation}
\sigma/\sigma_\mathrm{dc} =\mathscr{S} (\frac{\hbar \omega}{2 \pi k_B T},
\omega \tau_{\mathrm{qp}})
\end{equation}
and is thus described in terms of the single parameter
$\hbar/\tau_\mathrm{qp}= 2 \pi (k_B T)^2/k_B T_0$. The scaling function $\mathscr{S}$ can
be analytically evaluated. The optical response of a local Fermi
liquid shows on a log-log plot a characteristic non-Drude plateau (see
lines on Fig.~\ref{Fig_THzFLscaling}). The onset of the plateau on the
low-frequency side occurs at the thermal frequency $\hbar \omega = 2
\pi k_B T$. 

If also the impurity scattering is introduced by a
frequency-independent imaginary part $\Sigma \rightarrow \Sigma - i
\Gamma_{\mathrm{imp}}$, the optical conductivity, normalized to the dc
value becomes a two parameter function. On the low-temperature side,
the onset of the plateau ceases to scale with $T$ but rather saturates
at a temperature at which the zero-frequency scattering from the
interactions coincides with the impurity scattering.

As signs of a plateau were also found in our measurements, we attempted to fit
the Berthod \textit{et al.} theory to our $\sigma_1$ data. We found that the low frequency
part can be fit well with $Z \Gamma_{\mathrm{imp}}=3$~meV and
  $T_0=150$~K (see Fig.\ 4(b) of main paper). The onset of the plateau, on the other hand, appears at
  a too-high frequency to account for our data, see Fig.\ \ref{Fig_THzFLscaling}. The plateau that we
  observe is thus not a consequence of a FL frequency dependence of
  the self energy. Furthermore, its onset does not scale with temperature.

 We also note that $T_0$ was for a doped Mott insulator described
 within the dynamical mean-field theory found to be about
 10$T_\mathrm{FL}$.  Insisting on Berthod \textit{et al.} description, despite
 the obvious deviations, for CaRuO$_3$ we find that
 $T_0/10>>T_\mathrm{FL} $.  $T_\mathrm{FL}$ thus appears anomalously
 low also from this perspective.

In Fig.\ \ref{Fig_THzFLscaling} we also plot the frequency dependence for the NFL regime if one assumes a bare Drude response, where the frequency-independent scattering rate $1/\tau_{\textrm{D}} \sim a T^{3/2} + 1/\tau_0$ reproduces the temperature dependence of the dc resistivity ($\tau_0 =$ 1.3~ps). In the NFL temperature range, the data below 0.6~THz is described well by this simple model.

\subsection{Comparison with previous optical studies}

\begin{figure}[tb]
\includegraphics[width=0.95\linewidth]{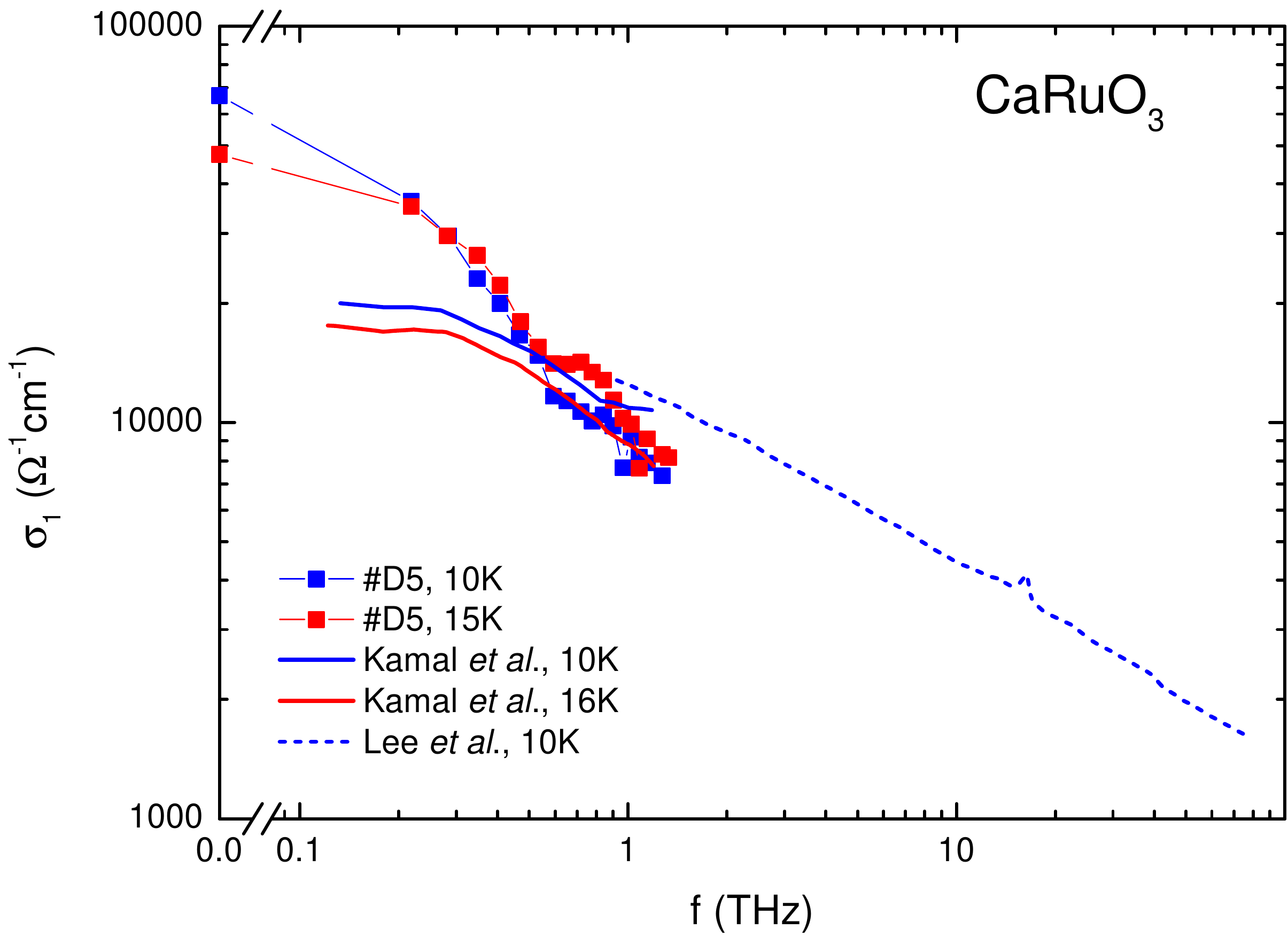}
\caption{\label{Fig_THzLiterature}Real part $\sigma_1$ of optical conductivity for our sample \#D5 at temperatures 10~K and 15~K. For comparison, THz data by Kamal \textit{et al}.\cite{Kamal2006} and infrared data by Lee \textit{et al} \cite{Lee2002} are displayed.}
\end{figure}

Our data obtained on sample \#D5 roughly compares to the results obtained previously by Kamal \textit{et al.} using THz time-domain spectroscopy \cite{Kamal2006}. In Fig.\ \ref{Fig_THzLiterature} we show our data at temperatures 10~K and 15~K, and data by Kamal \textit{et al.} for 10~K and 16~K. The main differences are that the low-frequency conductivity of our sample is much higher due to the improved sample growth, and that the order of the curves is exchanged: for our sample and above 0.3~THz, $\sigma_1$($T$=15~K) exceeds $\sigma_1$($T$=10~K), because here the sample is already in the relaxation regime, with excitation frequency larger than the optical scattering rate. Our data are also consistent with the previous infrared results by Lee \textit{et al.} \cite{Lee2002}, as shown with the dashed curve for 10~K in Fig.\ \ref{Fig_THzLiterature}. The slight mismatch in absolute conductivity we again attribute to the different sample quality.

\end{document}